\documentclass[onecolumn,authoryear]{els-mrw} 

\usepackage{amsmath,amssymb,amsfonts,amsthm,makeidx,graphicx}
\usepackage[rightcaption]{sidecap}
\usepackage{txfonts}
\usepackage{helvet}

\newcommand{\nat}{{Nature}}
\newcommand{\aap}{{A\&A}}
\newcommand{\aj}{{AJ}}
\newcommand{\apj}{{ApJ}}
\newcommand{\araa}{{ARA\&A}}
\newcommand{\apjl}{{ApJL}}
\newcommand{\apjs}{{ApJS}}

\newcommand{\mnras}{{MNRAS}}

\begin{document}

\chapter{Protoplanetary disk chemistry and structure}\label{chap1}

\author[1,2]{Merel L.R. van 't Hoff}%
\author[3]{Jennifer B. Bergner}%

\address[1]{\orgname{University of Michigan}, \orgdiv{Department of Astronomy}, \orgaddress{1085 S. University Ave., Ann Arbor, MI 48109-1107, USA}}
\address[2]{\orgname{Purdue University, Department of Physics and Astronomy, 525 Northwestern Avenue, West Lafayette, IN 47907, USA}}
\address[3]{\orgname{University of California, Berkeley}, \orgdiv{Department of Chemistry}, \orgaddress{Berkeley, CA 94720, USA}}


\maketitle

\begin{abstract}[Abstract]
Knowledge of the composition of material that will form planets is crucial to understand planetary diversity and the occurrence of potentially habitable planets. Ultimately, it is the chemistry in circumstellar disks that determines the global make up of planetary systems, as the dust in these disks grows into giant planet cores and rocky planets, the gas becomes incorporated in giant planet atmospheres, and the ices can be delivered to rocky planets by comets and meteorites. With the advent of ALMA a decade ago and the recent launch of JWST, the composition of the disk gas and ice can now be studied in great detail. This review will provide an overview of our current knowledge of the disk chemical structure, focusing on the six elements essential to life on Earth: carbon (C), hydrogen (H), nitrogen (N), oxygen (O), phosphorus (P) and sulfur (S). 
\end{abstract}

\begin{glossary}[Keywords] Astrochemistry, circumstellar disks, circumstellar gas, circumstellar dust, molecular gas, ice composition, small molecules, millimeter astronomy, submillimeter astronomy, infrared astronomy.
\end{glossary}

\begin{glossary}[Glossary]

\term{Chemical inheritance} Molecules formed during a certain stage of star formation become incorporated unaltered into the next stage. 

\term{Chemical reset} Molecules formed during a certain stage of star formation are destroyed or altered before becoming incorporated into the next stage. Also referred to as `processing'.

\term{CHNOPS} The six elements essential to life on Earth: carbon (C), hydrogen (H), nitrogen (N), oxygen (O), phosphorus (P) and sulfur (S). 

\term{Coma} Cloud of gas and dust produced from a comet's surface, caused by ice sublimation when the comet passes close to the sun.

\term{Complex organic molecule} In the context of astrochemisty, a carbon-bearing molecule containing at least six atoms. 

\term{Condensation} Phase change from gas to ice. Often referred to as `freeze out' or `depletion'. 

\term{Depletion} Generally, the removal of a molecular species from a certain phase, and is often used to describe the phase change from gas to ice (also referred to as `freeze out' or `condensation'). It is also more loosely used to indicate a lower molecular abundance in the gas beyond the `loss' expected from ice formation alone. 

\term{Desorption} Phase change from ice to gas. Also referred to as `sublimation'. 

\term{Deuterium} Heavy isotope of hydrogen containing a proton and a neutron in the nucleus.  

\term{Dissociation} Destruction of the bond between two atoms. 

\term{Endothermic reaction} Reaction that consumes energy. 

\term{Envelope} Remnant of the parent molecular cloud that shrouds a newly forming star.  Infall of the envelope results in accretion of material onto the disk and protostar.

\term{Exothermic reaction} Reaction that releases energy.

\term{Freeze out} Phase change from gas to ice. Also referred to as `condensation' or `depletion'. 

\term{Freeze-out temperature} Temperature at which a molecular species transitions from being primarily present in the gas phase to being primarily present in the ice. Often interchangeably used with `sublimation temperature'.

\term{Habitable} A planet that is potentially hospitable to life.  Generally implies a rocky/terrestrial planet composition with a temperate climate allowing for the existence of liquid water, and access to biogenic elements (CHNOPS).

\term{Herbig star} A young star on the high end of the stellar mass spectrum, more than two times as massive as the sun.

\term{Hydrogenation} The addition of an hydrogen (H) atom to a molecule.

\term{Ice line} Radius in the disk midplane where the temperature is equal to the freeze-out temperature of a molecular species. Inside its ice line, a molecular species is predominantly present in the gas phase, while outside its ice line it is predominantly present in the ice. Also referred to as `snowline'. 

\term{Interferometer} Array of correlated telescope antennae which are combined to produce images with high spatial resolution.

\term{Ion} Atom or molecule with a net electrical charge due to the addition or removal of elections, causing the positive charge of the nucleus and negative charge of the electrons to no longer balance out. Most ions in disks are positively charged. 

\term{Ionization} Process converting electrically neutral atoms or molecules into charged atoms or molecules (ions) through the addition or removal of electrons. In disk environments, electron removal is the dominant ionization process.  

\term{Isotopes} Different `versions' of an element due to different numbers of neutrons in their nuclei (e.g., $^{12}$C and $^{13}$C). 

\term{Isotope fractionation} Preferential enrichment or depletion of a particular isotope in a molecular reservoir relative to the cosmic isotope ratio. 

\term{Isotopologues} Different `versions' of a molecular species due to incorporation of different isotopes (e.g., $^{12}$C$^{16}$O, $^{13}$C$^{16}$O and $^{12}$C$^{18}$O). 

\term{Line emission/absorption} Discrete spectral feature due to a transition between two quantum states of an atom or molecule that involves the emission or absorption of a photon. 

\term{M dwarf} The lowest-mass class of stars (less than roughly half the mass of the sun).  M dwarfs are the most common type of star. 

\term{Midplane} Equatorial plane of the protoplanetary disk, characterized by the coldest and densest conditions.  Planet formation takes place within the midplane.

\term{Molecular layer} Chemically rich vertical layer in the disk that is sufficiently shielded from UV radiation to not destroy all molecules and warm enough to prevent molecules from being present only as ice. 

\term{Photoablation} The removal of atoms and/or small molecules from grain surfaces by energetic photons.

\term{Photodesorption} Phase change from ice to gas induced by UV or X-ray photons. 

\term{Photodissociation} Destruction of the bond between two atoms upon absorption of a UV or X-ray photon. 

\term{Radical} Atom, molecule, or ion that has at least one unpaired electron. 

\term{Refractory material} Material that remains solid up to high temperature (a few 100 to a few 1000 Kelvin, depending on the composition of the material). Typically referred to as `dust grains'. 

\term{Self-shielding} Ability of certain molecular species (e.g., H$_2$, CO, N$_2$, H$_2$O) to protect themselves from photodissociation as these molecules dissociate only upon absorbing photons of specific wavelengths. If a large enough column of these molecules is present, molecules in the top layer absorb all dissociating photons, protecting molecules at deeper layers.   

\term{Snowline} Radius in the disk midplane where the temperature is equal to the freeze-out temperature of a molecular species. Inside its snowline, a molecular species is predominantly present in the gas phase, while outside its snowline it is predominantly present in the ice. Also referred to as `ice line'. 

\term{Snow surface} Vertical height in the disk, varying with radius, where the temperature is equal to the freeze-out temperature of a molecular species. Above its snow surface, a molecular species is predominantly present in the gas phase, while below its snow surface it is predominantly present in the ice.

\term{Solids} Material in either refractory (dust grains) or ice form. 

\term{Speciation} The distribution of an element amongst chemical species.

\term{Sublimation} Phase change from ice to gas. Often referred to as `desorption'.  

\term{Sublimation temperature} Temperature at which a molecular species transitions from being primarily present in the ice to being primarily present in the gas. Often interchangeably used with `freeze-out temperature'. 

\term{T Tauri star} A young star on the low end of the stellar mass spectrum, less than two times the mass of the sun.

\term{Volatility} Tendency for a molecule to exist in the gas versus solid state, related to its binding energy. Generally treated as a spectrum from hypervolatiles (e.g.~CO, N$_2$, CH$_4$) to volatiles (CO$_2$, H$_2$O, CH$_3$OH) to (semi-)refractories (NH$_4^+$ salts, amorphous/mineral solids).

\end{glossary}

\begin{glossary}[Nomenclature]
\begin{tabular}{@{}lp{34pc}@{}}
ALMA & Atacama Large Millimeter/submillimeter Array \\ 
CHNOPS & Carbon, Hydrogen, Nitrogen, Oxygen, Phosphorus, Sulfur \\
IR & Infrared \\
ISM & Interstellar medium \\
JWST & James Webb Space Telescope \\
LTE & Local Thermodynamic Equilibrium \\
PDR & Photodissociation region \\
UV & Ultraviolet \\
\end{tabular}
\end{glossary}

\begin{glossary}[Learning objectives]
By the end of this chapter you will understand 
    \begin{itemize}
        \item What can be learned from molecular line observations
        \item How molecules in disks can be observed 
        \item The main processes important for the chemistry in disks
        \item The global chemical structure of disks
        \item What the main carriers are for the six elements essential to life on Earth
        \item What snowlines are and their relevance to planet formation 
         
    \end{itemize}
    
\end{glossary}


\section{Introduction: why study disk chemistry?}

\begin{figure}[!h]
\centering
\includegraphics[width=0.5\textwidth]{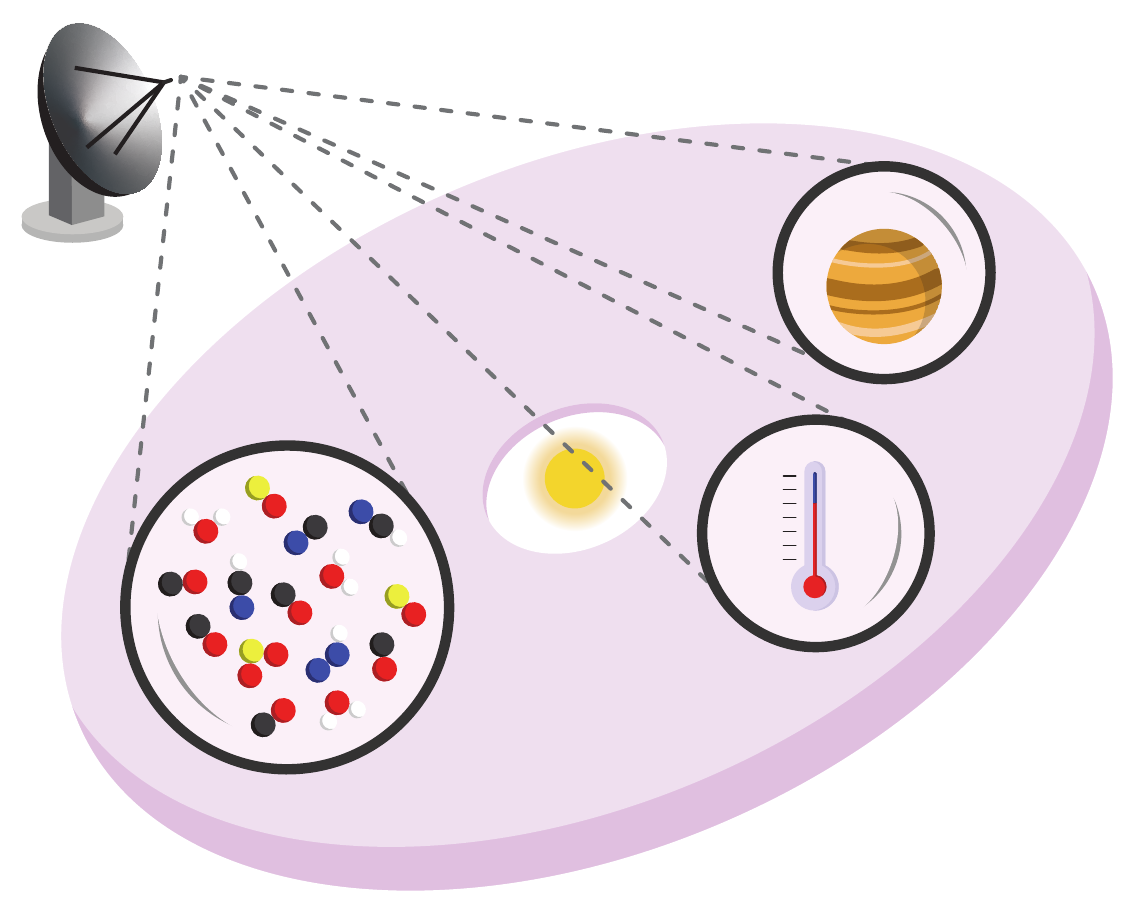}
\caption{Cartoon illustrating an example of different aspects of protoplanetary disks that can be studied with molecular line observations: the chemical content of the gas and ice that may be incorporated into planets, physical properties such as temperature, and disturbances in the disk velocity structure due to forming planets.  }
\label{fig:intro}
\end{figure}

With the detection of several thousand planets around stars other than our Sun in the last decades and the capability of JWST and (upcoming) ground-based facilities to study their atmospheres in great detail, we are entering an exciting era of characterizing planets and searching for Earth-like and potentially habitable planets. Constraining the origin of different types of planets and how common or rare they may be requires a detailed understanding of the chemical composition of the gas and solids from which planets form (Fig.~\ref{fig:intro}). For example, in the simplest view, which molecules are present in the gas at the formation location of a giant planet will determine the elemental composition of its atmosphere. In addition, the condensation (`freeze-out') of molecules onto dust grains may influence the growth of small grains into larger bodies, directly impacting the planet formation process itself. Furthermore, emission from molecules probes the physical conditions, such as temperature and ionization level, that are important to understand disk evolution and planet formation. Finally, molecular emission reveals the kinematics in the disk, providing a way to infer the presence of forming planets. Chemistry in protoplanetary disks has therefore been a topic of many observational and modeling studies.  This review aims to summarize the current state of the field at a level suitable to non-experts and early-career scientists.

\begin{SCfigure}[0.7][h]
\caption{Illustration of the three forms, or phases, atoms and molecules can take in disks. Refractory elements such as iron (Fe), magnesium (Mg) and silicon (Si) are almost exclusively present in dust grains (unless the temperature exceeds hundreds of Kelvin). These refractory grains also contain significant amounts of carbon (C) and oxygen (O). More volatile elements (CHNOPS) are either present as gas or ice on the surface of the dust grains, depending on the temperature (see Section~\ref{sec:RadialStructure}).}
\includegraphics[width=0.5\textwidth]{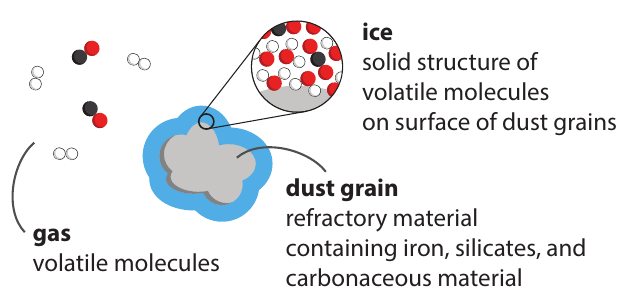}
\label{fig:dust-ice-gas}
\end{SCfigure}

We will here focus on the six elements essential to life on Earth: carbon (C), hydrogen (H), nitrogen (N), oxygen (O), phosphorus (P) and sulfur (S)-- often abbreviated as CHNOPS.  In the interstellar medium (ISM), these elements can exist either in a refractory form (solid phase unless the temperature exceeds a few hundred Kelvin, referred to as dust grains) or a volatile form (either gas or ice depending on the temperature) (Fig.~\ref{fig:dust-ice-gas}). The speciation of these elements into different chemical carriers within a protoplanetary disk profoundly impacts the outcomes of planet formation, and will be discussed in detail in this review. We will focus on the volatile states (gas and ice), as those can be probed directly with infrared and (sub)millimeter observations.  

It is important to recognize that the chemical makeup of a protoplanetary disk is, to some degree, influenced by the chemistry of earlier evolutionary stages.  The process of star and planet formation begins with a dense molecular cloud in which about half of the total available carbon, a substantial fraction of oxygen, and the majority of sulfur and phosphorus are present in refractory dust grains. Despite the relatively low densities (10$^4$--10$^5$ particles cm$^{-3}$) and low temperatures ($\sim$10 K), that make it hard for two atoms to find each other and react, molecules do form from the remaining volatile atoms. H$_2$ and CO are the most abundant molecules in the molecular cloud gas, but most material is frozen in an icy layer on the surface of the dust grains. Ice mantles are host to a rich chemistry leading to the production of simple hydrides (e.g.~H$_2$O, CH$_4$, NH$_3$), as well as CO$_2$ and so-called `complex' organic molecules: C-bearing molecules containing at least six atoms. During collapse of the molecular cloud to form a protostar, the ice and gas are subjected to heating and radiation processing, which leads to further chemical processing and increases the degree of chemical complexity present in both the gas and ice phase.  This protostellar material represents the initial chemical conditions for a protoplanetary disk.  The chemical state of a disk is likely a combination of inherited cloud/protostellar material, and material that undergoes further processing in the unique disk physical environment.  The balance between chemical inheritance and reprocessing, as well as the coupling between physics, dynamics, and chemistry, are both major themes in the study of protoplanetary disk chemistry.

This chapter is structured as follows. Section~\ref{sec:PhysicalStructure} starts with a brief overview of the disk physical structure, focusing on aspects directly impacting the chemistry. Section~\ref{sec:Chemistry} then introduces the dominant processes for gas-phase (Sect.~\ref{sec:GasChemistry}) and ice-phase (Sect.~\ref{sec:IceChemistry}) chemistry. The resulting disk chemical structure in vertical (Sect.~\ref{sec:VerticalStructure}) and radial (Sect.~\ref{sec:RadialStructure}) directions is detailed in Sect.~\ref{sec:ChemicalStructure}. Differences between young and mature disks are briefly described in Sect.~\ref{sec:YoungDisks}. A description of the main CHNOPS carriers in disks is given in Sect.~\ref{Sec:ElementalBudget}, and the chemistry for their isotopes is presented in Sect.~\ref{sec:Isotopes}. Sections~\ref{sec:COMs} and \ref{sec:Substructures}, respectively, detail our knowledge about (complex) organic molecules in disks and chemical substructures. Finally, a brief introduction to molecular line observations and spectroscopy at microwave (Sect.~\ref{sec:Microwave}) and infrared (Sect.~\ref{sec:Infrared}) wavelengths is presented in Sect.~\ref{sec:Observing}, and a succinct overview of analysis techniques and models is given in Sect.~\ref{sec:Analysis}. We end with a brief outlook for the future (Sect.~\ref{sec:Outlook}).


\section{Protoplanetary disk physical structure} \label{sec:PhysicalStructure}

\begin{figure}[!t]
\centering
\includegraphics[width=\textwidth]{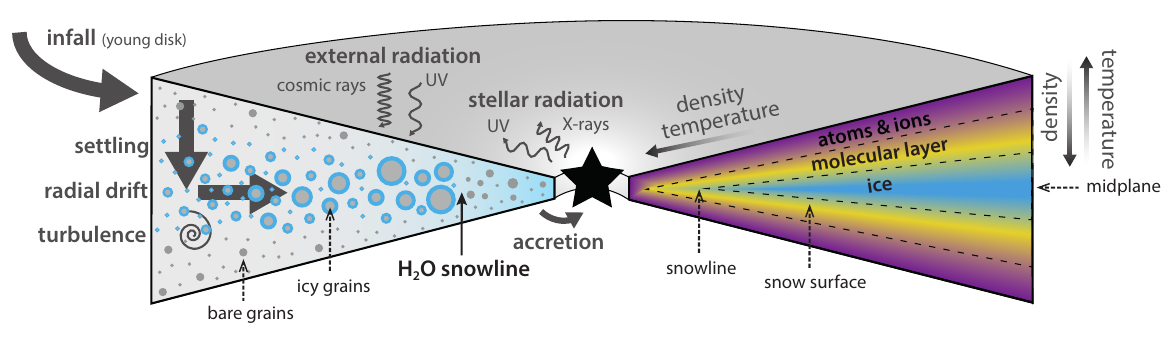}
\caption{Illustration of the physical and chemical structure of protoplanetary disks. Young embedded disks, also called protostellar disks, are expected to have a similar structure albeit warmer with less dust evolution (i.e., settling, drift and growth) and are embedded in remnants of the parent cloud, called the envelope. Infall of material on the disk is therefore most important for young disks.}
\label{fig:diskstructure}
\end{figure}

Protoplanetary disks are host to a wide range of physical conditions, spanning cold quiescent environments similar to a molecular cloud, to hot irradiated environments similar to a photodissociation region (PDR).  Overall, there are strong vertical and radial gradients in the physical structure.  The density increases moving from the disk surface to the midplane in the vertical direction, and decreases moving from the star towards the outer disk in the radial direction.  The temperature and radiation fields decrease from the disk surface to the midplane, and decrease from the inner to outer disk. A schematic of the physical structure is presented in Fig.~\ref{fig:diskstructure}, and a more detailed discussion is presented by e.g., \citet{Henning2013}.

\paragraph{Temperature}

The thermal structure of the disk is determined by the processing of stellar and interstellar radiation by dust particles. The main heating mechanism in the disk surface is called photoelectric heating, which proceeds as follows.  When a UV (ultraviolet) photon hits a dust grain, part of its energy can free an electron from the grain surface, while the remaining energy heats the grain itself.  The free electrons in turn interact with and heat the gas. In the region of the disk closest to the star, accretion heating (loss of gravitational energy as gas accretes onto the star) can also play a dominant role in heating the dust grains. As the accretion rate typically decreases with age, this process is expected to be most important in young disks that are still embedded in remnants of the parent cloud. Dust grains in the bulk of the disk are heated by infrared radiation from other grains. This way, heating of grains in the inner region and surface layers of the disk results in a cascade of absorption and re-emission of infrared radiation by dust grains throughout the disk. The gas, in turn, is heated through collisions with the dust. The main cooling mechanism for the disk is the emission of thermal radiation from the dust grains, which peaks in the infrared. In most of the disk, the density is high enough for collisions between gas and dust to be frequent enough that the gas temperature is equal to the dust temperature. This is not the case in the upper surface layers, where the gas and dust temperature are decoupled, and the gas temperature is determined by the balance of photoelectric heating versus cooling through atomic and molecular line emission.

\paragraph{Density}

The radial surface-density (density integrated over the full height of the disk) profile of disks can be approximated on large spatial scales by a power-law with an exponential fall off towards the outer edge.  It is typically assumed that the dust surface density is 100$\times$ smaller than the gas surface density based on interstellar gas-to-dust ratios, though it is contested whether the same ratio should hold in disks. Recently, high-resolution observations with ALMA have revealed that the small-scale dust distribution is not, in most cases, well described by a smoothly varying power-law surface density profile: the dust is often locally concentrated in rings or spiral structures, referred to as dust `sub-structures'.  Similarly, high-resolution observations have shown that disk gas molecules exhibit sub-structuring.  It remains unclear to which extent the bulk gas (H$_2$) is substructured or whether this is a chemical effect (see more in Section 8). The millimeter dust is generally more radially compact than the gas due to drift of large solids towards the star, though it is difficult to measure disk sizes precisely because of the vanishing emission at the cold and low-density outer edge. The vertical distribution of gas in disks is often considered to be set by hydrostatic equilibrium, where gravity and thermal pressure are in equilibrium, and the ratio of thermal to gravitational energy increases with disk radius.  While the small dust should remain well coupled to the gas, larger dust grains will `settle' to a more vertically compact region near the midplane.  In outer disk regions, gravity becomes too weak to confine the material to a thin plane and the thermal energy expands the disk in the vertical direction. This puffing up of the outer disk is called `flaring'.

\paragraph{Radiation}

In addition to heating the disk, the radiation field also plays a direct role in the chemistry, inducing processes like ionization and dissociation which drive subsequent reactions (Section~\ref{sec:Chemistry}).  Of particular importance are UV radiation, X-rays and cosmic rays. UV radiation is generated by the accretion shock from material falling onto the star and in the stellar photosphere. This type of radiation has the smallest penetration depth and is only present in the surface layers of the disk.  Small dust grains are the major source of UV opacity, and so the UV field is strongly dependent on the dust distribution. X-ray emission can penetrate deeper into the disk and young stars show strong X-ray emission due to their enhanced stellar and magnetic activity compared to more evolved stars. Cosmic rays have the largest penetration depth and can provide a source of ionization in the disk midplane. The amount of cosmic rays present in disks is still uncertain: on the one hand stellar winds may reduce the flux of galactic cosmic ray particles impinging onto the disk. On the other hand, cosmic rays may be created locally through the acceleration of particles at shock surfaces (see e.g., the discussion by \citealt{Zhang2024}).  

\paragraph{Dust dynamics}

As already hinted at, dynamics plays an important role in setting the physical structure of a disk. In the vertical direction, large grains tend to gravitationally settle towards the midplane, while turbulence can stir material from the midplane back to the surface layers.  In the radial direction, a phenomenon called `radial drift' acts on intermediate-sized dust grains (often called `pebbles'): as the gas orbits at sub-Keplerian velocities due to gas pressure support, intermediate-size grains experience a headwind. Due to this gas drag, they lose angular momentum and spiral towards the region of maximum pressure. In a smooth disk this means that pebbles drift into the star, though the presence of local pressure maxima (pressure `bumps') can halt the drift of pebbles at disk substructures.  Smaller grains are well coupled to the gas and do not experience radial drift. Larger solids follow Keplerian orbits and are not meaningfully influenced by the presence of the slower-moving gas. The exact grain size that is susceptible to gas drag depends on many factors and varies as a function of disk radius, but is typically around millimeter--centimeter scale. The ongoing growth and fragmentation of grains throughout the disk lifetime means that settling, turbulence, and drift are operating on a constantly evolving population of dust grains.  These dynamical processes are greatly important to disk chemistry, since they impact the physical conditions (radiation field, temperature, ionization) and also result in transport of volatiles between different disk regions.  Lastly, the disk physical conditions can also evolve due to the astrophysical context of star formation: for example, disks are expected to cool as accretion onto the star slows down and the envelope dissipates, while short periods of enhanced accretion (so-called `accretion bursts') result in temporary increases in temperature.

\section{Astrochemical processes} \label{sec:Chemistry}

\begin{figure}[!t]
\centering
\includegraphics[width=\textwidth]{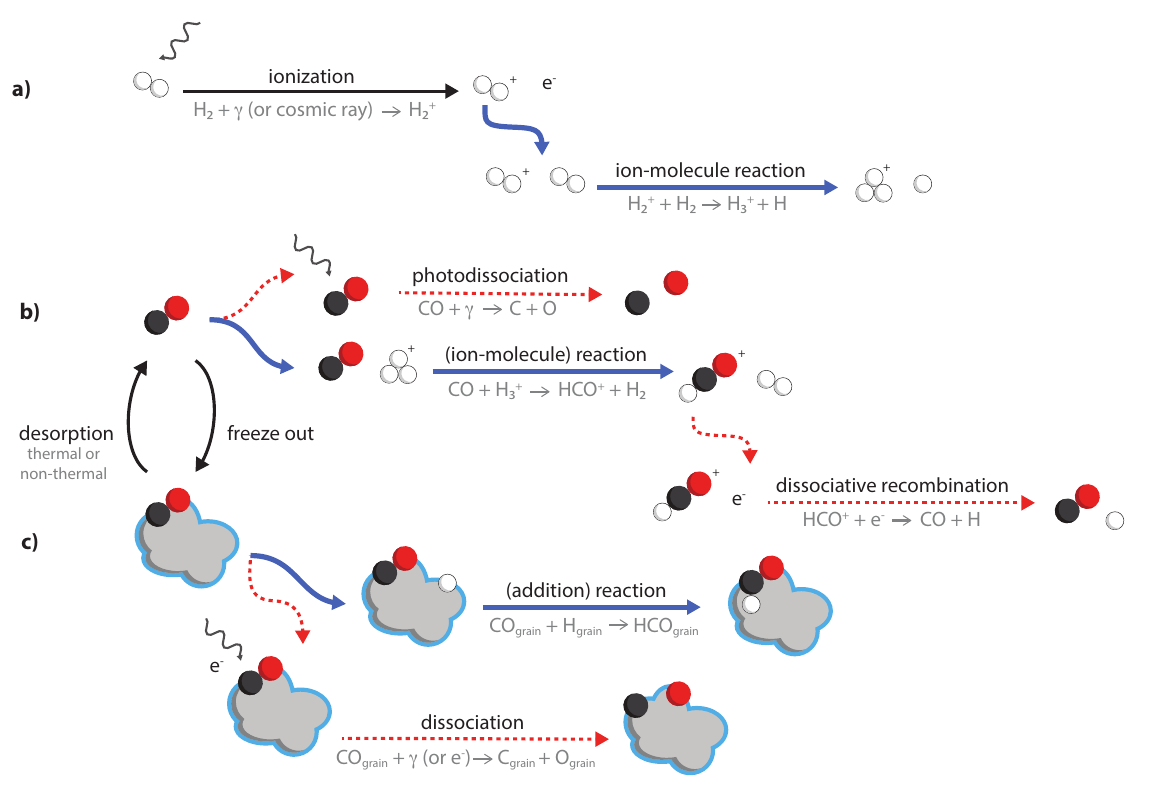}
\caption{Illustration of the main chemical reaction types in disks. The dominant ionization pathway (a) results in the formation of the H$_3^+$ ion. Examples of different types of gas-phase (b) and ice-phase reactions (c) are shown for CO, but also apply to other molecules. Routes that lead to the formation of new molecules and potentially greater chemical complexity are shown in blue, and destruction routes are shown in red. Photons are denoted with $\gamma$ in the reactions.}
\label{fig:reactions}
\end{figure}

Throughout much of the disk, the temperatures and densities are too low for the gas chemistry to reach steady-state, that is, a state in which the molecular abundances are set by the current physical conditions and do not change over time (as long as the physical conditions do not change). Instead, molecular abundances depend on the relative rates of different chemical reactions and change during the lifetime of the disk, even if the physical conditions remain constant. As reaction rates depend on physical conditions, the molecular abundances at any time depend on the physical conditions at current and past times. Infrared surveys of large populations of disks suggest a median dust-disk lifetime of several million years, indicating that chemical processes must operate on shorter timescales to be relevant in disks, although the lifetime of the gas disk is more uncertain. Reaction pathways and rates are typically derived from laboratory experiments or quantum-physical calculations.  Reactions can take place either in the gas phase or within the icy mantles coating grain surfaces (Fig.~\ref{fig:reactions}).  Some molecules have formation pathways both in gas and ice phases (e.g.~formaldehyde, H$_2$CO), while other molecules are thought to form in only one phase (e.g., methanol, CH$_3$OH, can only form on a grain surface).

\subsection{Gas-phase chemistry} \label{sec:GasChemistry}

\subsubsection{Bulk of the disk}

At the low temperatures ($<$ 100 K) present in the bulk of the disk, reactions may proceed only if they (i) release energy (`exothermic' reactions) and (ii) do not have an activation energy barrier. At the same time, the low-to-intermediate densities ($\lesssim 10^8$ cm$^{-3}$) favor so-called two-body reactions which do not require a third colliding species to dissipate reaction energy. Reactions between two neutral molecules are typically endothermic and/or have an activation barrier, whereas the transfer of a proton from an ion to a neutral molecule is usually exothermic and barrierless. Ion-neutral chemistry therefore dominates in the bulk of the disk gas. 

\paragraph{Ion-neutral reactions}

To drive ion-neutral reactions, and thus gas-phase chemistry, a source of ionization (UV radiation, X-rays, cosmic rays or radionuclides) is needed. An ionization event typically leads to H$_3^+$ formation since hydrogen is the most abundant element. H$_3^+$ then powers a rapid ion-molecule chemistry which can lead to the production of complex molecular ions (Fig.~\ref{fig:reactions}). For example, H$_3^+$ can transfer a proton to CH$_2$ to form CH$_3^+$. This molecular ion can then react with HCN, forming CH$_3$CNH$^+$. 

\paragraph{Dissocative recombination reactions}

Neutral molecules can be formed through a process called dissociative recombination, in which a free electron recombines with a molecular ion which then dissociates into two stable molecules or a molecule and atom.  This can produce both simple and relatively large neutral molecules. For example, CH$_3$CNH$^+$ can recombine with an electron to form CH$_3$CN and a hydrogen atom. 

\paragraph{Other reaction types}

In addition to driving the formation of molecules, radiation also leads to their destruction, as molecular bonds can be broken upon absorption of UV photons in a process called photodissociation. While other gas-phase molecule formation pathways, such as neutral-neutral reactions (reaction between two neutral molecules, producing two different neutral species) and radiative association (reaction between two atoms or molecules forming a larger molecule while emitting a photon), are possible, they are in most cases much less efficient than ion-molecule chemistry.

\subsubsection{Inner disk}

At the high temperatures ($\sim$100 -- 5000 K) and densities ($\gtrsim 10^{12}$ cm$^{-3}$) in the inner few au of the disk, the chemistry is fast enough to approach equilibrium.  Even so, dynamical processes may prevent equilibrium from being reached. In the absence of intense sources of ionizing radiation, neutral-neutral reactions with barriers $\gtrsim$ 100 -- 1000 K start playing an important role, and 3-body reactions become important in regions with densities $>10^{12}$ cm$^{-3}$. Molecules remain abundant in the gas phase until the temperature exceeds $\sim$2500 -- 3500 K at the inner edge of the disk, where they are destroyed by thermal dissociation. 

\subsection{Ice chemistry}  \label{sec:IceChemistry}

\subsubsection{Transitions between gas and ice}

A major factor governing the chemical structure of disks is the partioning of molecules between the gas phase and the frozen/solid phase, called the `ice' phase.  A given molecule's tendency to exist in the gas or solid phase, referred to as its `volatility', is determined by its binding energy: the strength of interaction between the molecule and a solid surface. In the case of interstellar ice molecules, the surface is typically an amorphous H$_2$O-dominated ice layer (that is, the molecules are arranged in a disordered rather than crystalline state) which coats a refractory dust grain.  A `freeze-out temperature' is often used to approximate the temperature at which a molecule transitions from primarily gas-phase to primarily ice-phase (though this balance formally depends on the density as well).  Molecules with lower binding energies have lower freeze-out temperatures, i.e.~remain in the gas phase down to lower temperatures. `Freeze-out' and `condensation' are usually used synonymously to describe the removal of gas-phase molecules onto dust grains. The term `depletion' is also used sometimes, but this is not recommended due to the more ambiguous meaning. Similarly, `sublimation' and `desorption' both describe the ice to gas transition.

Ice-phase molecules can be desorbed into the gas phase either by thermal desorption, if the grain is warmer than the molecule's sublimation temperature, or via non-thermal desorption mechanisms.  Photodesorption refers to desorption induced by photons like UV or X-rays, and cosmic rays can also cause ice desorption.  Typically, only small molecules can be photodesorbed intact, whereas larger molecules are photodesorbed as fragments. Excess energy released upon the formation of a molecule on the grain surface may also result in its sublimation, called chemical desorption or reactive desorption. 

\subsubsection{Reactions in the ice}

Molecules can still participate in active chemistry after freezing out onto grains (Fig.~\ref{fig:reactions}).  In fact, ice-phase chemistry is more efficient than gas phase chemistry for certain reaction classes, including the formation of many organic molecules.  Importantly, the ice mantle can dissipate excess reaction energy (acting as a `third body'), stabilizing products which would otherwise dissociate into fragments in the gas phase.  This explains why some molecules like H$_2$ and CH$_3$OH cannot form in the gas phase, but can form on grain surfaces.  Ices also concentrate potential reactants and enable multiple reaction attempts for barriered reactions.  

New molecules can form in the ice through either `energetic' or `non-energetic' pathways.  Energetic processing refers to chemical processes induced by photons, electrons, or cosmic rays, which dissociate stable molecules into highly reactive radical species (atom or molecule with at least one unpaired electron).  These radicals can recombine barrierlessly with neighboring radicals to form larger molecules, or if the grain temperature is high enough, radicals can diffuse over the surface to encounter new reaction partners. Non-energetic chemistry is driven by atom addition and abstraction reactions, in which atomic species (mainly H, but also potentially O) are added to or removed from a stable molecule.  In both cases, the result is the formation of reactive radical species that can, once again, recombine with other radicals to form larger molecules.  While most neutral-neutral reactions cannot be thermally activated at temperatures relevant to ices ($\lesssim$150 K), a notable exception is the formation of ammonium salts, which can occur at a few tens of K.


\section{Protoplanetary disk chemical structure} \label{sec:ChemicalStructure}

\subsection{Vertical structure} \label{sec:VerticalStructure}

The strong temperature and density gradients together with differences in the radiation fields at various disk locations create a three-layered chemical structure in the vertical direction (Fig.~\ref{fig:diskstructure}; \citep{AikawaHerbst1999}).  

\paragraph{The surface layer} 

\begin{SCfigure}[][!t]
\centering
\includegraphics[width=0.7\textwidth]{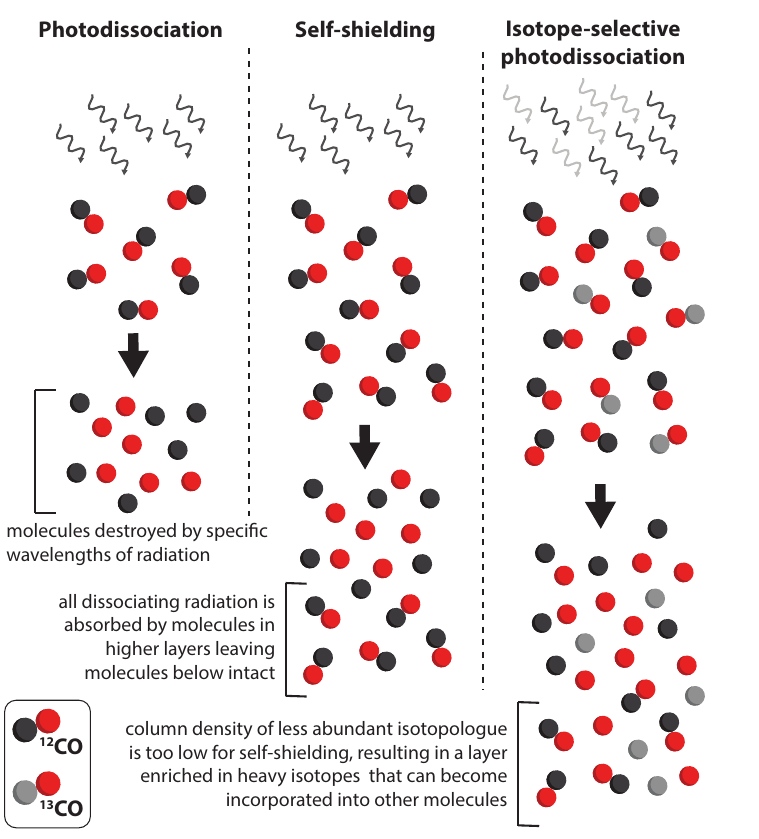}
\caption{Illustration of photodissociation by line emission (left), where molecules are destroyed by radiation of specific wavelengths, self-shielding (middle), where the top layer of molecules absorbs all dissociating radiation protecting the molecules in deeper layers, and isotope-selective photodissociation (right), where the less abundant isotopologue has a column density too low for self-shielding, resulting in a layer enriched in the rare isotope that can then become incorporated into other molecules.}
\label{fig:photodissociation}
\end{SCfigure}

In the disk surface layers, the chemistry is dominated by photoprocesses. The stellar UV radiation and the interstellar radiation field ionize and dissociate molecules. The disk surface therefore has parallels with regions in the dense interstellar medium that are irradiated by massive stars, so-called photodissociation regions (PDRs). Molecules that are dissociated by line radiation (i.e., dissociated by light at specific frequencies), for example, H$_2$, CO and N$_2$, can self-shield at sufficiently high column densities. This means that all radiation at the specific dissociative frequencies gets absorbed by molecules closest to the disk surface, protecting molecules deeper in the disk (Fig.~\ref{fig:photodissociation}). In addition, as photons with certain frequencies can dissociate both H$_2$ and CO (or H$_2$ and N$_2$), H$_2$ can partially shield CO and N$_2$. Water (H$_2$O) is another molecule present in the disk surface layer because, in addition to self-shielding, it can form efficiently via neutral-neutral reactions when the gas temperature is higher than 400 K. X-ray photons can penetrate deeper into the disk than UV photons and ionize helium, producing He$^+$ which is able to destroy tightly bound molecules like CO, therefore driving a rapid hydrocarbon chemistry.

\paragraph{The warm molecular layer}

Underneath the disk surface is a warm molecular layer that is sufficiently shielded from UV irradiation such that not all molecules are destroyed. The survival of molecules leads to a rich chemistry that is driven by X-rays. As the temperature in this layer is $\sim$30--70 K, the least volatile molecules such as H$_2$O and NH$_3$ will be frozen onto the dust grains. The freeze-out of water, and consequently the removal of a large fraction of the oxygen budget from the gas phase, results in carbon-to-oxygen (C/O) ratios elevated above the disk overall value, leading to a carbon-dominated gas chemistry  (more in Sect. \ref{sec:CandO}). Furthermore, these lukewarm temperatures allow for grain surface chemistry resulting in, for example, formation of formaldehyde (H$_2$CO). In less opaque regions (e.g., in the outer disk), UV photons may drive photodesorption of ice species. As the UV opacity is dominated by small dust grains, grain growth and settling allows ionizing and dissociating photons to penetrate deeper into the disk. The vertical extent of the warm molecular layer is thus strongly coupled to the evolution of the dust.  

\paragraph{The midplane} 

The deep interior of the disk is completely shielded from UV and X-ray radiation, causing the temperature to drop below 20 K.  Freeze-out of molecules and hydrogenation reactions on grain surfaces dominate the chemistry. Because of its low binding energy to grain surfaces, H$_2$ will remain in the gas even in the midplane, but most other molecules are expected to be frozen out.  Ice-phase chemistry may still proceed in the midplane to some extent if cosmic rays are present, though it is generally thought to be rather chemically inactive. \\

\begin{figure}[!t]
\centering
\includegraphics[width=0.9\textwidth]{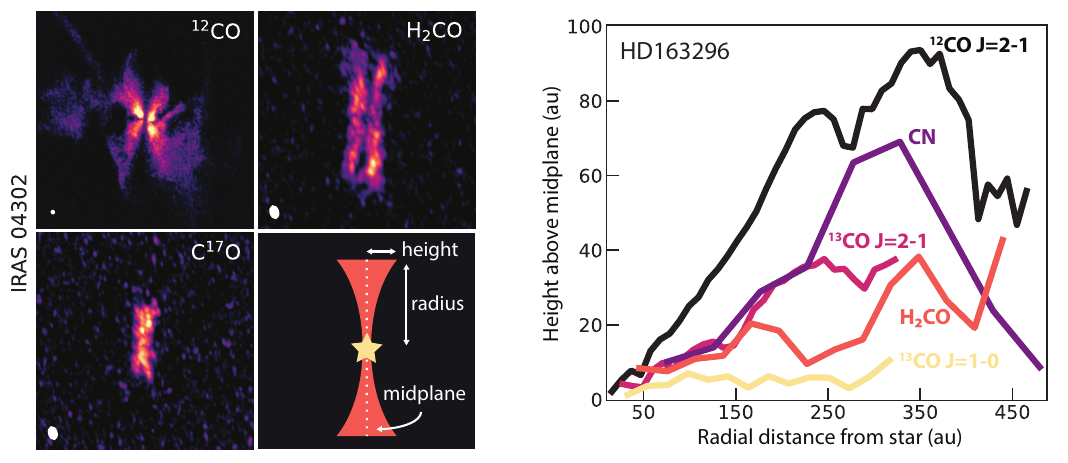}
\caption{Examples of vertical stratification in disks. The vertical structure can be probed directly in edge-on disks, as shown for the young disk IRAS 04302 (left; ALMA observations originally presented by \citealt{vantHoff2020} and \citealt{Lin2023}). In less inclined disks such as HD163296, the emitting height can be derived from observations with high spectral and spatial resolution (right; adapted from \citealt{Paneque-Carreno2023}; \copyright ESO).}
\label{fig:verticalstructure}
\end{figure}

\vspace{-0.4cm}
Most of the molecular line emission from disks thus originates from the chemically-rich warm molecular layer or near the disk midplane inside that molecule's snowline. The exact emitting region for a molecular transition will depend on the chemistry that determines where a molecule is most abundant, and on excitation, as higher-energy transitions will be excited in warmer gas than lower-energy transitions. In edge-on disks, vertical stratification of different molecules and transitions can be observed directly, while for more face-on disks the emitting height of molecular emission can be inferred from observations with high spatial and spectral resolution (Fig.~\ref{fig:verticalstructure}).

\subsection{Radial structure and snowlines} \label{sec:RadialStructure}

In the disk midplane, molecules will freeze onto dust grains when the temperature drops below a given molecule's freeze-out temperature. The midplane radius where this happens is referred to as the radial `snowline' or `ice line'. Each molecule has a unique snowline location determined by its freeze-out temperature. Combined with the radial temperature gradient in the disk, this results in sequential freeze out of different molecules, with the least volatile molecules freezing out closest to the star (Fig.~\ref{fig:radialstructure}). For example, the H$_2$O snowline is located at temperatures of $\sim$150 K, corresponding to a few au for T Tauri stars, while CO does not freeze out until the temperature drops below $\sim$20--30 K.  Moving vertically away from the midplane, a freeze-out transition will occur at increasingly larger radii since the same temperature is achieved further from the star.  This leads to a two-dimensional `snow surface' which reflects the ice-gas transition for a given molecule at all radii and elevations in the disk (Fig.~\ref{fig:diskstructure}).  The positions of snow lines will vary from disk-to-disk depending on, for example, the stellar mass, and will also change as the disk temperature structure evolves during the disk lifetime.

\subsubsection{Deriving snowline locations}

\begin{figure}[!h]
\centering
\includegraphics[width=0.9\textwidth]{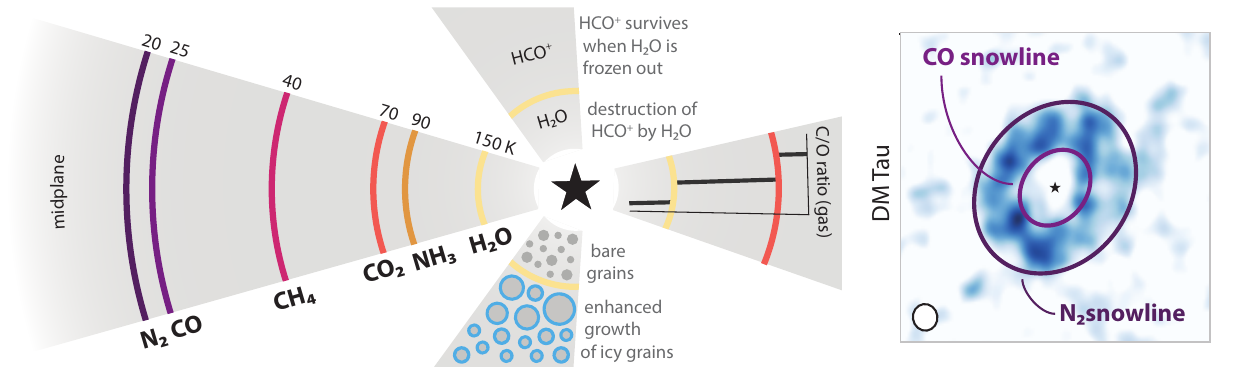}
\caption{Illustration of snowlines. Left: Schematic of different snowline locations in the disk midplane, and examples of their effects on the disk chemical and physical structure. The left sector shows the snowline locations of the main carbon, oxygen and nitrogen carriers. Indicated freeze-out temperatures are approximate and depend on the density and the composition of the existing ice layer. The bottom sector dipslays the effects of the water snowline on the dust grains. The right sector presents a simplified view of how the carbon-to-oxygen (C/O) ratio of the gas changes across the H$_2$O and CO$_2$ snowlines (based on \citealt{Oberg2011}). The top sector illustrates how chemistry can be used to locate snowlines, as depicted for the H$_2$O snowline. Right: The CO and N$_2$ snowlines are traced by the inner and outer edge, respectively, of a bright N$_2$H$^+$ emission ring as N$_2$H$^+$ is formed from gaseous N$_2$ and destroyed by gaseous CO (adapted from \citealt{Qi2019} with permission; \copyright AAS). The ellipse in the lower left corner marks the synthesized beam-size of the interferometric observations, indicating the spatial resolution. }
\label{fig:radialstructure}
\end{figure}

It is often difficult to measure snowline locations directly. For example, a depletion of CO in the outer disk midplane is hard to observe (except when the disk is viewed edge-on) because gas-phase CO remains abundant in higher vertical layers. Therefore, chemical effects have been used instead to estimate snowline locations. For example, N$_2$H$^+$ forms through the reaction between N$_2$ and H$_3^+$, and is primarily destroyed by CO. N$_2$H$^+$ is therefore abundant where CO is absent from the gas (beyond the CO snowline) and where N$_2$ is present in the gas (inside the N$_2$ snowline). Because the N$_2$ snowline is slightly further out than the CO snowline, this results in ring-shaped N$_2$H$^+$ emission where the inner radius is roughly associated with the CO snowline (Fig.~\ref{fig:radialstructure}). Similarly, as gas-phase H$_2$O is the main destroyer of HCO$^+$ in the disk midplane, HCO$^+$ can be used to trace where H$_2$O has frozen out. 

\subsubsection{Importance of snowlines}

Snowlines have significant consequences for the chemistry of planet formation.  First, molecule freeze-out significantly lowers the available molecules for gas-phase reactions, while increasing the inventory of molecules available for ice-phase chemistry.  Since different types of reactions occur in the gas versus ice, snowlines will affect what new chemistry may occur at a given disk location.  Snowlines also alter the total abundance of volatile elements in the gas versus ice phase.  For instance, the freeze-out of abundant species like H$_2$O, CO$_2$, and CO can significantly increase the total amount of oxygen in the ice (and decrease the amount in the gas).  This will in turn impact the composition of planetary cores and atmospheres that form from the disk solids and disk gas, respectively, in a particular disk location.  The influence of disk snowlines on the elemental ratios of planet-forming material is posited as a way to connect observable properties of exoplanets (e.g.~atmospheric C/O ratios) to their formation origins \citep{Oberg2011}.  Snowlines will also regulate what molecules are present in icy planetesimals (like modern-day asteroids and comets), and therefore what volatile inventory could potentially be delivered via impacts to planetary surfaces.

Snowlines are also important for planet formation physics.  For instance, freeze-out of certain molecules may increase the stickiness of ice-coated grains compared to bare grains, increasing the efficiency of dust growth and in turn promoting the first steps of planet formation.  Just interior to snowlines, slight increases in the gas pressure can also cause local pile-ups of dust, again promoting particle growth.

\subsection{Young disks} \label{sec:YoungDisks}

The chemistry of young disks that are still surrounded by remnants of the envelope ($\lesssim0.5-1$ Myr old) have not yet been studied in as much detail as those of more-evolved mature protoplanetary disks. While the overall chemical structure as described above is expected to be present, there are already some notable differences observed between young protostellar disks and mature protoplanetary disks. For example, young disks are typically warmer due to higher mass accretion rates and the presence of an envelope, which results in the snowlines being at larger radii from the star. In particular, young disks smaller than $\lesssim$100 au appear to have no CO ice, with CO freezing out only in the envelope beyond the disk \citep{vantHoff2020}. Another topic of recent research is whether the accretion of material from the envelope onto the young disk can create a shock, and whether this has a chemical signature \citep[e.g.,][]{Sakai2014}. The sulfur-bearing species SO and SO$_2$ have been postulated as good tracers of such accretion shocks, and while SO emission has been observed around the disk--envelope interface for a few systems, how uniquely these molecules are associated with a shocked region remains undetermined.


\section{Elemental budget} \label{Sec:ElementalBudget}

Tracking the volatile elements H, C, O, N, S, and P is of particular interest due to their universal importance in terrestrial biochemistry.  In planet-forming environments, these elements can be speciated into different molecular carriers, which may be partitioned into either the solid or gas phase depending on their volatility.  For instance, at temperatures around 50 K, carbon stored in CO will exist as a gas, while carbon stored in CO$_2$ will exist as a solid.  The incorporation of the biogenic elements into planets therefore involves (i) the chemical processes that set the relative abundances of different molecular carriers, and (ii) the sublimation behavior which regulates the gas versus ice partitioning of each molecule.  

\subsection{Hydrogen}

Hydrogen makes up the vast majority of interstellar material (74\% of the entire mass budget).  In dense star-forming regions (including disks), most H is in the form of H$_2$, which is quite chemically inert. Hydrogen is incorporated into molecules early in the star-formation sequence, resulting in the formation of hydrides like H$_2$O, NH$_3$, CH$_4$, CH$_3$OH, and other organics. In ices, these molecules account for much of the H budget, because due to its high volatility, H$_2$ will not be retained in the solid state. 

The bulk of the disk mass is thus composed of H$_2$, and H$_2$ is also expected to dominate the composition of primordial planetary atmospheres. Because H$_2$ has no permanent dipole moment, it is not observable in cold environments (see more in Sect.~\ref{sec:Microwave}). This means that the mass of a disk, and therefore the amount of material available to form planets, cannot be measured directly. The best alternative tracer of the gas mass is the deuterated form of H$_2$, HD (see more in Sect.~\ref{sec:Deuterium}). However, suitable transitions fall in the far-IR and are not observable with current facilities. Gas masses are therefore often estimated either from the dust mass obtained from the continuum emission assuming a gas-to-dust ratio of 100, or from optically thin CO isotopologue emission adopting interstellar values for the isotope ratios and the CO abundance with respect to hydrogen (10$^{-4}$).

\subsection{Carbon \& Oxygen} \label{sec:CandO}
  
The main volatile carbon carriers in protoplanetary disks are CO and CO$_2$, while O is mainly partitioned into H$_2$O, CO$_2$, and CO. Other C- and O-bearing volatiles are low in abundance compared to CO, CO$_2$, and H$_2$O, and therefore less important to the bulk elemental compositions of planet-forming gas and solids.  Roughly 50\% of the cosmic C budget, and a few tens of percent of the O budget, are expected to be stored in a refractory form, i.e.~minerals, amorphous carbon or macromolecular organics.

In the cool, outer regions of protoplanetary disks, CO and H$_2$O appear to be `depleted' from the gas in the observable surface layers of the disk, by factors of 10--100 compared to the interstellar budgets \citep[e.g.,][]{Ansdell2016,Du2017}.  CO$_2$ is not directly observable in cool disk gas due to having no rotational transitions (see Sect. \ref{sec:Microwave}).  This depletion is thought to reflect the coupled chemistry and dynamics within the disk: as ice-covered grains grow in size, they gravitationally settle and bring icy material (dominated by H$_2$O and CO) to the midplane.  This volatile material may be prevented from re-entering the gas by some combination of chemical reprocessing into less volatile material, physical sequestration in large solid bodies, and drift inwards towards the star.  The net effect is that oxygen and, to a lesser extent carbon, are removed from the gas in elevated layers of the disk.  Moreover, grain growth removes small dust from the disk atmosphere, allowing more UV radiation to penetrate into the molecular layer of the disk and driving photochemical processes.  

Other chemical indicators tell us that oxygen must be more depleted than carbon: the relatively high gas-phase abundances of hydrocarbons (C$_2$H, cyclic-C$_3$H$_2$) and nitriles (CN, HCN, HC$_3$N, CH$_3$CN) necessitate a high carbon-to-oxygen ratio in the gas \citep[e.g.,][]{Bergin2016,LeGal2019}.  One way to increase the gas C/O ratio is photodissociation of CO, the main gas-phase carbon carrier, followed by H$_2$O formation from the free oxygen.  However, this process can only achieve C/O ratios as high as unity, whereas ratios around 1.5--2 are implied by the bright hydrocarbon emission.  Photoablation of carbon grains in the UV-irradiated disk surface has been suggested as an additional source of carbon to achieve C/O ratios $>$1 \citep{Bosman2021}. While CO does not appear to be depleted in young disks, nondetections of water emission suggest that the abundance is low \citep{Harsono2020}. Depletion of water may thus already occur during the embedded phase.

Volatile C and O that is sequestered on icy pebbles in the midplane is likely subject to further transport in the disk.  As mentioned in Section \ref{sec:PhysicalStructure}, pebble drift causes solids to spiral inwards towards the star, except when drift is halted by the presence of substructures (pressure `bumps').  There is mounting evidence that, for systems with unhindered drift (small, smooth disks), the inner disk is strongly enriched in volatile gas, interpreted as a signature from the drift and sublimation of icy material originating in the outer disk \citep[e.g.,][]{Banzatti2023}.  On the other hand, for systems with pressure bumps (large, sub-structured disks), it is expected that inner disks should be relatively dry and that O- and C-rich ices are accumulated at the substructure locations \citep[e.g.,][]{Vlasblom2024}.  Understanding radial ice transport will be critical to inferring the reservoir of volatiles available to planets forming at different distances from the host star in systems with different physical substructures.
  
\vspace{-0.1cm}
\subsection{Nitrogen}
Constraints on the total disk nitrogen budget are more limited.  N$_2$ is suspected to be the dominant volatile N carrier, but it is undetectable as a homonuclear diatomic (see Section~\ref{sec:Microwave}).  Observations of trace N carriers must be used instead to follow the nitrogen chemistry-- in the gas phase, this includes protonated N$_2$ (N$_2$H$^+$), nitriles (CN, HCN, HC$_3$N, CH$_3$CN), and in very few cases ammonia (NH$_3$).  Disks appear to be especially efficient sites for nitrile formation, and HCN is commonly detected both in the cool outer disk and the hot inner disk.  Currently, there is no evidence that chemical--dynamical evolution affects the disk nitrogen budget as it does the C and O budget.  One possible explanation is that the major N carrier, N$_2$, is quite volatile and chemically inert, meaning that it is not susceptible to sequestration on grains either by freeze-out or by processing to a less volatile carrier.  There are currently few constraints on the ice-phase N budget in disks, though some clues come from solar system comets.  Low cometary nitrogen abundances have long puzzled the community, though in recent years it has emerged that much of the solid N budget may be in the form of semi-refractory NH$_4^+$ salts, which form through thermal reactions between NH$_3$ and ice-phase acids like H$_2$S, HCOOOH, and HCN \citep{Poch2020}. Early JWST observations of protoplanetary disk ices show hints of ammonia-H$_2$O hydrates and OCN$^-$, but as of 2024 there have been no claims of NH$_4^+$ nor other trace N carriers like HCN \citep[e.g.,][]{Sturm2023}.  

\vspace{-0.1cm}
\subsection{Sulfur}
Several gas-phase sulfur carriers have been detected in disks: CS, SO, H$_2$S, H$_2$CS, and SO$_2$. However, these gas-phase carriers still represent just a small fraction of the cosmic S budget \citep{Keyte2024}, which is also the case for protostellar systems.  Of the (trace) gas-phase carriers, there are hints that a surprisingly high fraction may be in an organic form like H$_2$CS. Much of the S reservoir is expected to be locked in refractory grains, an idea supported by relatively high S abundances in stellar photospheres which reflect freshly accreted solids \citep{Kama2019}.  While OCS ice has been tentatively found in JWST disk observations \citep{Sturm2023}, the total S inventory in ice appears also insufficient to explain the full S budget, implying a larger contribution of mineral forms of sulfur.  Interestingly, while H$_2$S is the dominant ice-phase S reservoir in comets, it has yet to be detected in the ice phase either in disks or in earlier stages of star formation.   

\vspace{-0.1cm}
\subsection{Phosphorus}
The phosphorus budget in protoplanetary disks is virtually unconstrained: no detections of gas or solid-phase phosphorus carriers have been made.  In prior star forming stages, gas-phase P carriers are only detected in shocked regions, implying that most of the P reservoir is locked in refractory minerals \citep{Bergner2022}.  Given the non-detection of any gas phase P tracer in disks, it is likely that most P is similarly stored in refractory grains.  In the solar system, the phosphorus carrier PO was detected in the coma of comet 67P \citep{Rivilla2020}, so a trace volatile P component may also be present but remaining undetected in other systems.

\section{Isotopes} \label{sec:Isotopes}

Isotope fractionation is the preferential enrichment or depletion of a particular isotope in a molecular reservoir relative to the cosmic isotope ratio. Isotopes are different `versions' of an element due to different numbers of neutrons in their nuclei. For example, the main isotope of carbon is $^{12}$C, with $^{13}$C being a factor $\sim$70 less abundant in the local ISM (Wilson \& Rood 1994). Molecules that incorporate different isotopes are called isotopologues (e.g., CO isotopologues include $^{12}$C$^{16}$O, $^{13}$C$^{16}$O, $^{12}$C$^{18}$O). Fractionation is sensitive to the physical conditions in which a molecule formed, such as temperature, radiation field, and ionization rate. Molecular isotope ratios are therefore diagnostic of the origin of molecules within interstellar settings. For example, fractionation is often particularly effective at low temperatures due to the small, but non-zero, zero point energy. This is the mimimum energy a molecule can have based on quantum mechanical principles. The zero point energy depends on the mass of the molecule, and is lower for an isotopically heavier molecule (e.g., $^{13}$CO) compared to an isotopically lighter variant ($^{12}$CO). The incorporation of an heavier isotope into a molecule can thus result in the release of energy, while the reverse reaction requires energy. At temperatures below the zero point energy difference, enrichment with the heavier isotope is therefore favored. Fractionation signatures observed in protoplanetary disks may be inherited from prior stages of star formation, or reflect in situ chemical processes.

\vspace{-0.1cm}
\subsection{Deuterium} \label{sec:Deuterium}

The less abundant stable isotope of hydrogen is called deuterium and instead of the $^2$H notation as used for other elements, this isotope is denoted D. Deuterium was formed in the big bang and has a cosmic abundance of roughly $10^{-5}$ with respect to hydrogen. Deuterium fractionation is initiated by the exchange reaction H$_3^+$ + HD $\leftrightarrow$ H$_2$D$^+$ + H$_2$, where the backward reaction is slightly endothermic (i.e., requiring energy) and therefore the forward direction is heavily favored at low temperatures.  Deuterium can then be transferred from H$_2$D$^+$ to other molecular carriers, resulting in enhanced molecular D/H ratios. Due to the relatively large mass difference between H and D, the difference in zero point energy results in very strong fractionation effects in cold star-forming environments, with molecular D/H ratios up to a few thousand times higher than the cosmic D/H ratio. Deuterium fractionation can also operate at somewhat higher temperatures ($>$30 K), via CH$_3^+$ + HD $\leftrightarrow$ CH$_2$D$^+$ + H$_2$. Both low-temperature and high-temperature fractionation pathways are inferred to be at play in the outer and inner disk regions, respectively. Deuterated molecules detected in protoplanetary disks are DCO$^+$, DCN, N$_2$D$^+$, and C$_2$D.

While there are no direct constraints on ice-phase deuteration levels within protoplanetary disks, high deuteration levels are seen in cometary ices in the modern-day solar system.  In particular, the high D$_2$O/HDO ratio in comets cannot be explained by in situ disk chemistry, and is inferred to reflect the inheritance of pre-stellar icy material.  Indeed, D/H is enhanced relative to the solar D/H in most rocky and icy solar system bodies, but to varying degrees, suggesting a balance between inherited and reprocessed reservoirs of solid hydrogen carriers (e.g.~H$_2$O, organics) during the assembly of planetesimals. D/H ratios in the solar system giant planets are close to the solar D/H ratio, implying that the parent disk gas did not experience significant fractionation.  

\vspace{-0.1cm}
\subsection{Oxygen and Nitrogen isotopes} 

The main oxygen and nitrogen isotopes are $^{16}$O and $^{14}$N, respectively, with $^{18}$O, $^{17}$O, and $^{15}$N less abundant in the local ISM compared to the main isotopes by factors of $\sim$560, $\sim$1790, and $\sim$450, respectively \citep{Wilson1994}. Due to their higher mass, the relative mass difference between different isotopes is small, and kinetic chemical effects at low temperatures, as observed for H and D, are minimal. Instead, oxygen and nitrogen fractionation primarily proceed via a photodissociation mechanism.  The major gas-phase carriers CO and N$_2$ undergo photodissociation via discrete line transitions and can therefore self-shield (see Sect.~\ref{sec:VerticalStructure} and Fig.~\ref{fig:photodissociation}). Rare isotopologues are less abundant, and therefore a longer path length is required before the dissociating radiation is fully absorbed.  There exist regions where main isotopologues are already self-shielding, while minor isotopologues are efficiently dissociated, freeing the rarer isotope for incorporation into other molecules (Fig.~\ref{fig:photodissociation}).  For instance, $^{14}$N$^{14}$N will self-shield closer to the edge of a cloud compared to $^{15}$N$^{14}$N, resulting in enrichment of $^{15}$N within photochemical products like CN and HCN.  Similarly, CO will self-shield before C$^{18}$O, enhancing the production of H$_2^{18}$O over H$_2^{16}$O.  Photodissociation fractionation is expected to be important in disks given the strong UV field in the surface layers.  There is evidence for an increasing HC$^{14}$N/HC$^{15}$N ratio moving from the inner to outer region of a protoplanetary disk, consistent with enhanced photodissociation in the less-shielded outer disk \citep{Hily-Blant2019}.  $^{18}$O  enrichment in primitive solar system solids is similarly thought to reflect isotope-dependent photodissociation, though it remains debated whether the fractionation occurs during the disk stage or is inherited from the molecular cloud.  

\vspace{-0.1cm}
\subsection{Carbon isotopes}

Carbon may be fractionated either via a photodissociation mechanism, or a low-temperature exchange mechanism.  CO is the main gas-phase carbon carrier and, as noted above, is subject to self-shielding effects which should enrich photochemically produced molecules with $^{13}$C relative to $^{12}$C.  The slightly endothermic exchange reaction CO + $^{13}$C$^+$ $\leftrightarrow$ $^{13}$CO + C$^+$ is also expected to operate at temperatures $<$40 K, but requires high ionization rates to be an efficient fractionation pathway.  Observations of carbon fractionation in protoplanetary disks hint at the existence of two distinct carbon isotope reservoirs: hydrocarbons (e.g.~C$_2$H) appear enhanced in $^{12}$C, and CO in $^{13}$C \citep{Yoshida2022, Bergin2024}.  Several early measurements of fractionation in exoplanet atmospheres indicate strong $^{13}$C enrichments, which would require the accretion of $^{13}$C-enriched gas from the disk.  On the other hand, there are no strong signatures of carbon fractionation in the solar system.  $^{13}$CO$_2$ has been detected in both disk ices and hot inner-disk gas with JWST \citep{Grant2023, Sturm2023}, but due to the high optical depth of $^{12}$CO$_2$, extracting isotope ratios will be challenging.  Additional measurements of carbon fractionation in different disk gas reservoirs as well as exoplanet atmospheres are needed to provide more clarity on the importance of carbon isotope fractionation within disks.


\section{(Complex) organic molecules} \label{sec:COMs}

There is great interest in understanding the inventory of organic material available to seed forming planetesimals and planets with potential building blocks for prebiotic chemistry.  Given the energetic conditions associated with active planet formation, molecules are not expected to survive direct incorporation into planets; instead, delivering molecules intact to terrestrial planets likely requires post-formation impacts of icy planetesimals. In planet-forming environments it is thus the organic content of the ice that matters for terrestrial planets, but given the difficulties identifying complex molecules in ices (see Section~\ref{sec:Observing}), gas-phase studies are crucial to uncover the full organic reservoir.

\subsection{Outer disk gas} 

At millimeter wavelengths, the small organics HCN, C$_2$H, and H$_2$CO are commonly detected in protoplanetary disk gas. HCN and C$_2$H are thought to form primarily through in situ gas-phase chemistry.  The abundances of these species are well correlated in disks \citep{Bergner2019}, likely reflecting that both are photochemically produced in C-rich gas.  H$_2$CO may form on grains in the outer disk beyond the CO snowline, or via warm gas-phase chemistry interior to the CO snowline.  The total inventory of small organics in disks is inferred to be quite large, equivalent to tens to hundreds of Earth oceans per disk \citep{Oberg2021}. 

The larger molecules c-C$_3$H$_2$ (cyclic cyclopropynelidene), HC$_3$N (cyanoacetylene), and CH$_3$CN (methyl cyanide) are also commonly detected in cool disk gas \citep{Ilee2021}.  Meanwhile, CH$_3$OH (methanol), the most abundant organic in the preceding protostellar stage, has proved surprisingly challenging to detect in disks \citep{Walsh2016}.  Indeed, CH$_3$CN/CH$_3$OH ratios in disk gas are $\sim$2 orders of magnitude higher than protostellar values, implying a dramatic shift in the organic chemistry during the disk stage.  This is likely another `symptom' of the phenomenon discussed in Section~\ref{sec:CandO}: grain growth and disk dynamics sequester O-rich ices in the midplane, leaving behind an O-poor gas exposed to high levels of UV radiation.  CH$_3$CN and HC$_3$N are shown to form efficiently via photochemistry operating in O-poor gas, and may therefore serve as signposts for the general reshaping of the disk volatile reservoir \citep{Calahan2023}.  The efficient synthesis of nitriles during the disk stage is intriguing due to the particular importance of nitriles within prebiotic chemical reaction networks.  Complex nitrile emission appears to originate close to the disk midplane, supporting that these molecules are available near the site of planet formation.

\begin{figure}[!h]
\centering
\includegraphics[width=0.9\textwidth]{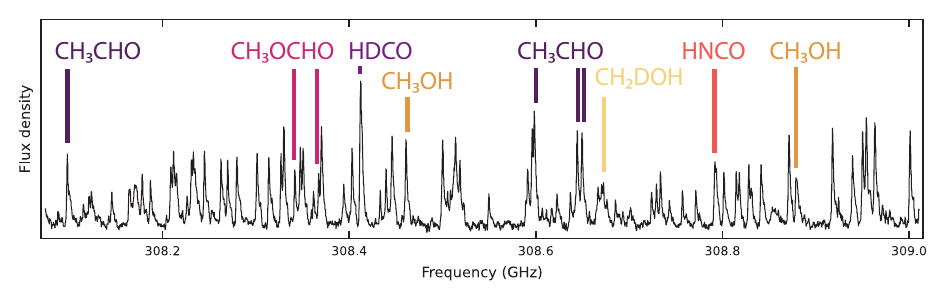}
\caption{Spectrum toward the disk around the outbursting young star V883 Ori displaying a wealth of emission lines from complex organic molecules. Several of the observed lines are annotated as an example. Data from the ALMA Large Program COMPASS (PI: Jes J{\o}rgensen). }
\label{fig:V883}
\end{figure}

In recent years, the detection of larger, O-bearing organics in disks has been enabled by focusing on young, outbursting disks as well as more massive Herbig systems.  The disk V883 Ori is an archetypical outbursting system: the disk has been transiently heated by an accretion burst, pushing the H$_2$O snowline out to $\sim$80 au (compared to $\sim$1 au in a typical T Tauri disk).  A rich array of complex organics (e.g.,~acetaldehyde: CH$_3$CHO, methyl formate: CH$_3$OCHO, acetone: CH$_3$COCH$_3$) have been detected in the gas of the V883 Ori disk (Fig.~\ref{fig:V883}; \citealt{Lee2019}), implying that organics are abundant in disks but are typically hidden in the icy midplane.  Meanwhile, disks around Herbig stars are warmer than T Tauri disks, and complex organics (e.g.~methanol: CH$_3$OH, dimethyl ether: CH$_3$OCH$_3$, ethylene oxide: c-H$_2$COCH$_2$), have now been detected towards several such systems \citep[e.g.,][]{Booth2021}. Importantly, these disks are expected to be too warm for CO to freeze out; since ice-phase CO is typically a requisite ingredient for CH$_3$OH formation, the detection of CH$_3$OH towards these systems necessitates inheritance of icy material into the disk.  

\subsection{Inner disk gas} 

Some small organics and hydrocarbons are detected at IR wavelengths tracing hot inner-disk gas from within the water snowline: HCN, C$_2$H$_2$, and CH$_4$.  From the Spitzer era, it was found that HCN/C$_2$H$_2$ ratios are systematically lower in disks around lower-mass M dwarf systems compared to solar-like systems \citep{Pascucci2013}.  There are hints that the nitrile/hydrocarbon ratios are also lower in the outer disks of M dwarf systems.  Possible explanations include the distinct radiation field of M dwarf versus T Tauri stars, or differences in the efficiency of icy pebble drift depending on disk mass. JWST has recently detected a remarkable complex chemistry in the inner disk around very low mass M dwarf stars, including large hydrocarbons (ethylene: C$_2$H$_4$, benzene: C$_6$H$_6$) as well as the nitrile HC$_3$N \citep{Tabone2023}.  This chemistry requires high C/O ratios and may reflect efficient destruction of carbon-rich grains in the inner disk \citep{Li2021}, and/or the trapping of O-rich icy pebbles further out in the disk that leaves the inner disk O-poor \citep[e.g.,][]{Mah2024}.  The importance of the so-called `soot line' (location where carbon-rich grains are thermally destroyed) to disk organic chemistry should become clearer with improved demographics of inner-disk chemistry with JWST.

\subsection{Ices} 

\begin{figure}[!h]
\centering
\includegraphics[width=0.9\textwidth]{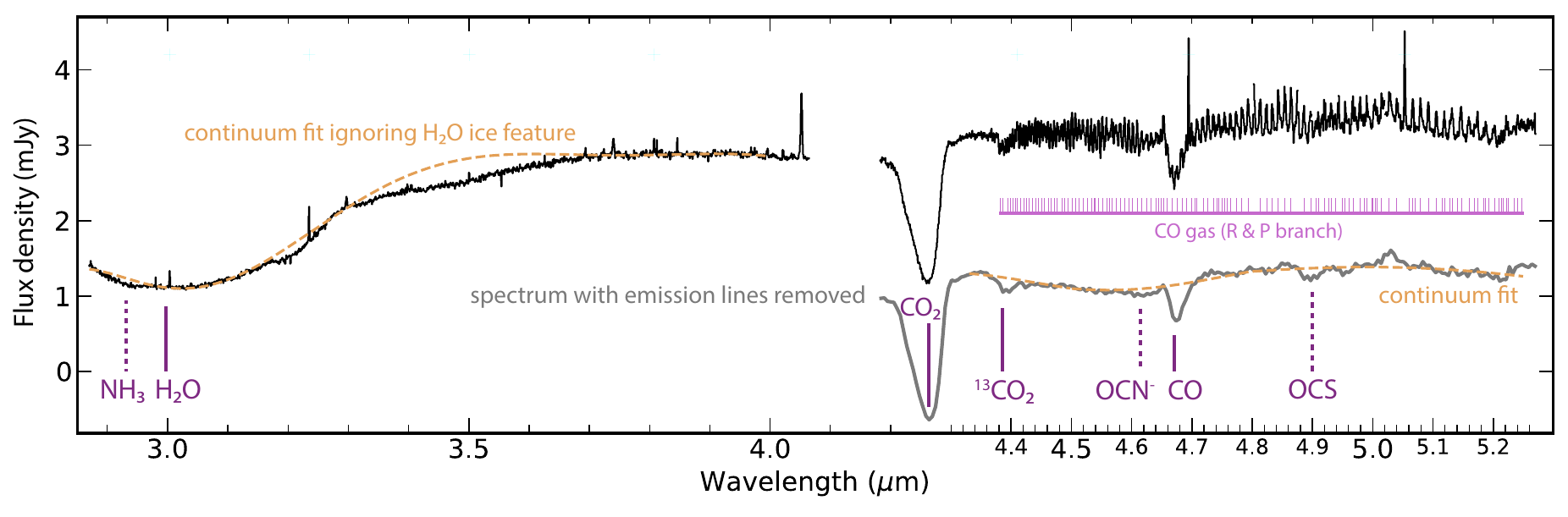}
\caption{JWST/NIRSpec spectrum of the edge-on disk HH 48 NE from the Early Release Science program Ice Age (black). The spectrum between 4.2 and 5.3 $\mu$m after removal of the gas emission lines is shown with a vertical offset (gray). The CO ro-vibrational emission lines are marked with light purple lines, and the ice absorption features with dark purple lines. Dotted lines indicate tentative detections. A local continuum fit excluding the CO$_2$ and broad H$_2$O ice features is shown with an orange dashed lines to highlight the tentative NH$_3$, OCN$^-$ and OCS ice features. Adapted from \citet{Sturm2023}; \copyright ESO.  } 
\label{fig:JWSTice}
\end{figure}

Organic molecules have not yet been detected in the ice-phase of protoplanetary disks, though the first JWST disk ice spectrum revealed hints of OCN$^-$ and OCS (Fig.~\ref{fig:JWSTice}; \citealt{Sturm2023}).  Based on protostellar ice measurements and disks with active ice sublimation (e.g.~V883 Ori), disk ices are expected to carry a wealth of organic complexity.  Indeed, solar system comets host an incredibly diverse assortment of organic molecules, including biologically relevant molecules like glycine (the simplest amino acid).  Meteorites contain significant quantities of macromolecular organic matter, which likely began as icy material that was subject to heavy radiation/thermal processing. The extent to which cometary and meteoritic organics reflect inherited prestellar/protostellar chemistry versus in situ disk ice processing remains an open question.  The agreement between the relative abundances of organics in the inner region of protostellar envelopes (called `hot corinos' for low-mass protostars and `hot cores' for high-mass protostars), outbursting protostellar disks, and comets is suggestive that the ice composition from early star-forming stages may be inherited relatively intact to the stage of icy planetesimal formation \citep{Drozdovskaya2019}.  Additional ice-phase organic molecule formation and/or reprocessing may also occur during the disk stage, particularly if icy grains are efficiently transported to moderately elevated, UV-rich layers of the disk, or in cases with high midplane ionization levels.

\section{Chemical substructure} \label{sec:Substructures}

\begin{figure}[!h]
\centering
\includegraphics[width=0.8\textwidth]{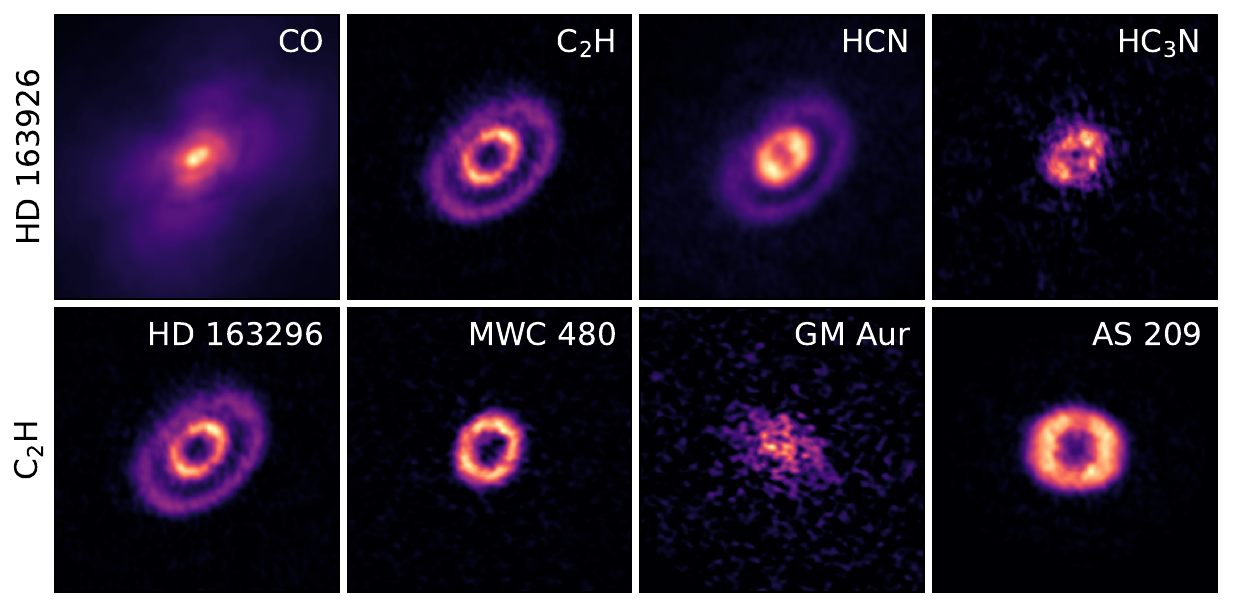}
\caption{Illustration of diversity in chemical substructure seen towards protoplanetary disks, made using data from the MAPS ALMA large program \citep{Oberg2021}.  Top: Different molecules imaged towards the HD 163296 disk.  Bottom: C$_2$H imaged towards different disks.}
\label{fig:chem_substructures}
\end{figure}

Since the advent of ALMA, it has become possible to spatially resolve the emission from protoplanetary disks at (sub-)millimeter wavelengths.  This led to the discovery that the dust is not smoothly distributed in most disks, but often exhibits rings, gaps, and spirals-- so-called `substructures'.  While the origin of these substructures is not definitively known, they are commonly thought to be associated with sites of active planet/planetesimal formation.  More recently, rings and gaps have been found to be ubiquitous for gas molecules as well (Fig.~\ref{fig:chem_substructures}; \citealt{Oberg2021}).  Chemical substructures show a remarkable diversity, with a range of morphologies observed both for different molecules imaged within the same disk, and for the same molecule imaged in different disks.  

The origins of chemical substructures remain murky.  Snowlines (CO, N$_2$) have largely been ruled out as an explanation.  In some cases, gas substructures are found to either correlate or anti-correlate with the presence of dust substructures, but there is no universal trend \citep{Law2021}.  Because chemistry is sensitive to a variety of factors (e.g.~temperature, radiation field, gas density), it may be that these driving factors combine in various ways to produce a range of outcomes for different molecules in different systems.  Identifying trends more definitively will require high-resolution chemical observations towards numerous additional disks.  Ultimately, this is a topic of pressing importance to predicting planetary compositions, since chemical substructuring means that planets forming in different disk locations could accrete quite different gas compositions.  Moreover, if there is indeed an association between dust and gas substructure, then the very process of planet formation may be reshaping the chemical composition in the neighborhood of the forming planet.

\begin{figure}[!h]
\centering
\includegraphics[width=0.7\textwidth]{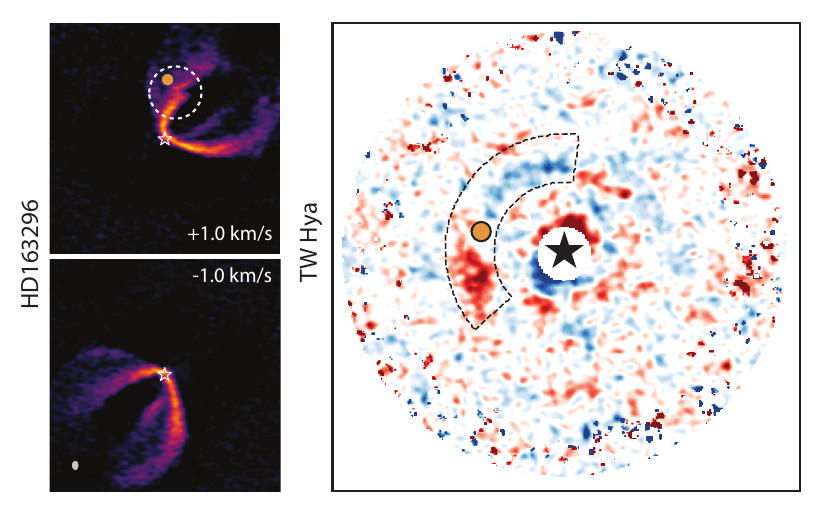}
\caption{Illustration of different techniques to infer the presence of planets through deviations in the velocity pattern using molecular line observations. The velocity deviations are marked by dashed regions and the inferred location of the planets are marked by orange circles. Left: A localized disturbance in the Keplerian pattern (`kink') of CO emission is visible in the northwestern side of the HD 163296 disk at redshifted velocities around 1.0 km/s (dashed contour in top panel), while a normal Keplerian pattern is visible at the same blueshifted velocities arising from the southeastern part of the disk (bottom panel). Figure adapted with permission from \citet{Pinte2018}; \copyright AAS. Right: Deviations from Keplerian velocity (dashed contour) in TW Hya stand out as strong residuals when the expected Keplerian velocity structure of the disk is subtracted from the observed velocity map for CS. Figure adapted from \citet{Teague2018}; \copyright AAS.}
\label{fig:substructures2}
\end{figure}

In addition to large-scale gas substructures, the high spatial and spectral resolution of ALMA has allowed the discovery of localized structures in individual frequency channels toward a few disks (Fig.~\ref{fig:substructures2}, left panels). These so-called `kinks' are believed to be due to (forming) planets that locally perturb the gas. Such perturbations can also be seen in the overall velocity profile of the disk as small deviations from Keplerian rotation (Fig.~\ref{fig:substructures2}, right panel). While these types of studies require very high signal-to-noise data, they may be one of the best ways to detect young planets embedded in disks. 

\section{Observing molecules in disks} \label{sec:Observing}

\begin{SCfigure}[][h]
\centering
\includegraphics[width=0.4\textwidth]{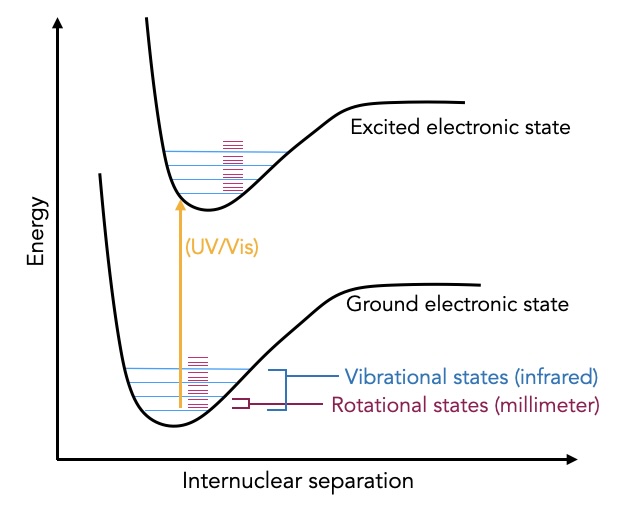}
\caption{Summary of molecular energy levels.  Each electronic state contains numerous vibrational states, which in turn contain many rotational states.  The large energy spacing between electronic states corresponds to transitions in the UV/visible, whereas vibrational transitions typically occur in the infrared, and rotational transitions in the millimeter.}
\label{fig:spectroscopy}
\end{SCfigure}

Molecules in disks, and astronomical environments in general, can be observed through their interaction with electromagnetic radiation.  A given molecule can only emit or absorb radiation at particular frequencies corresponding to energy differences between two quantized energy levels. Molecules have quantized electronic states, vibrational states and rotational states. As depicted in Figure \ref{fig:spectroscopy}, each electronic state contains many vibrational states, each of which contains many rotational states. Observations of molecules in disks rely predominantly on transitions between rotational and vibrational states: electronic transitions have wide spacings between energy levels, and are typically not populated at the relatively low temperatures ($\lesssim$1000 K) typical of protoplanetary disks. For most astrophysically relevant molecules, rotational transitions lie in the microwave (centimeter, millimeter and submillimeter wavelengths), and vibrational transitions in the infrared. Current (sub)millimeter facilities commonly used to observe disks are ALMA (the Atacama Large Millimeter/submilliter Array), NOEMA (the NOrthern Extended Millimeter Array) and the SMA (the Submillimeter Array). Infrared observations are best done from space, which is currently only possible with JWST (the James Webb Space Telescope), but some wavelength ranges are also accessible from the ground, with for example the VLT (Very Large Telescope) and the future ELTs (Extremely Large Telescopes). 

\subsection{Microwave observations} \label{sec:Microwave}

The lowest rotational levels have energies ($E_u$/$k_B$, where $k_B$ is the Boltzmann constant) as low as a few Kelvin, and are therefore readily excited at temperatures of only $\sim$10 K. This makes rotational spectroscopy the best tool for probing emission from cold interstellar environments.  Millimeter observations are also exceptionally sensitive and can detect molecules with low abundances down to $\sim10^{-11}$ with respect to H$_2$.  One limitation is that rotational transitions can only be observed for gas-phase molecules, as the interactions between molecules in ice lattices hamper rotations. Rotational transitions are also only possible for molecules that possess a permanent electric dipole moment. The electric dipole moment is a measure of the imbalance of charge within a molecule. Such an imbalance is required to produce an electric field that can couple to photons. Homonuclear molecules such as H$_2$ and symmetric molecules like CO$_2$ 
do not have a permanent dipole moment and can therefore not be observed via pure rotational transitions.

\begin{figure}[h]
\centering
\includegraphics[width=0.9\textwidth]{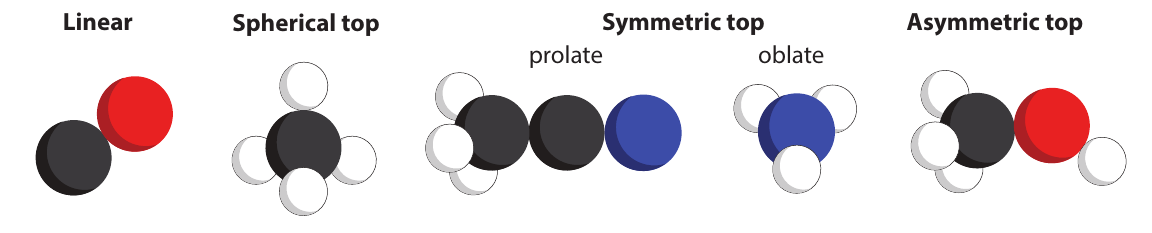}
\caption{Illustration of the four different molecular geometries or rotors: linear (e.g., CO), spherical top (e.g., CH$_4$), symmetric top, which can be either prolate (e.g., CH$_3$CH) or oblate (e.g., NH$_3$), and asymmetric tops, which are all molecules that do not fall into the previous three categories (e.g., CH$_3$OH). }
\label{fig:rotationalsymmetries}
\end{figure}

\begin{SCfigure}[][!h]
\centering
\includegraphics[width=0.6\textwidth]{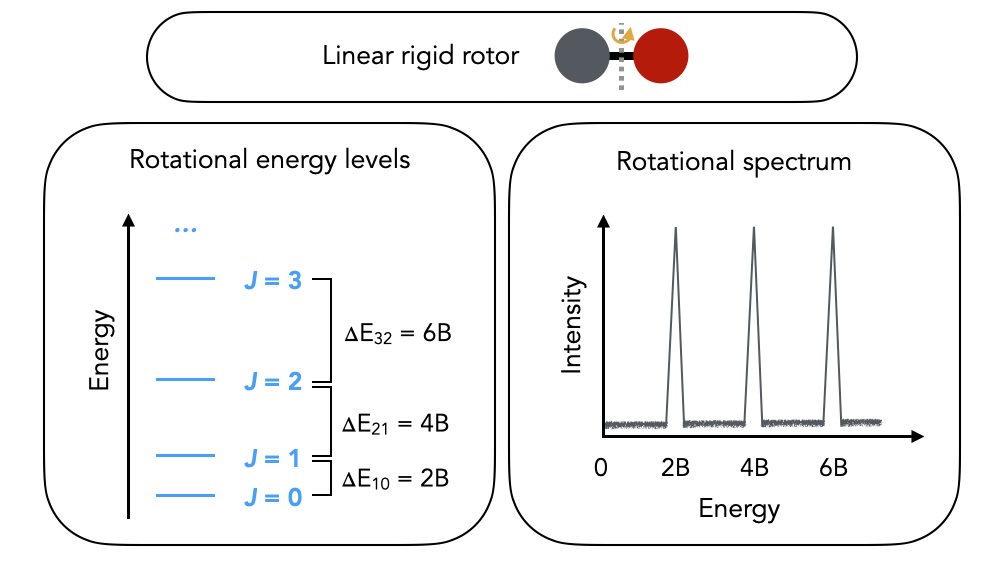}
\caption{Illustration of rotational energy levels for a linear rigid rotor.  Energy levels are given by $BJ(J+1)$, leading to transition energies $\Delta E$ of $2B(J+1)$ (left panel).  There is then a constant spacing of $2B$ between rotational transitions (right panel).}
\label{fig:spectroscopy_rot}
\end{SCfigure}

The energy spacing between different rotational states depends on the molecular geometry. Every molecule can be classified as one of four general types of rotor, determined by the moment of inertia along each of three orthogonal spatial axes (Fig.~\ref{fig:rotationalsymmetries}). The simplest case is for a linear molecule (e.g.~H$_2$, CO, HCN), where one moment of inertia is equal to zero. For linear molecules, rotation can therefore be treated as a `rigid rotor' (Fig.~\ref{fig:spectroscopy_rot}). Here, the rotational energy levels are given by $E = BJ(J+1)$, where $J$ is the rotational quantum number and $B$ the rotational constant, which is inversely proportional to the moment of inertia of the molecule.  
Rotational transitions are only possible between adjacent rotational states i.e.~$\Delta J = \pm1$.  The energy of a rotational transition, $\Delta E = E(J+1) - E(J)$, is then always equal to $2B(J+1)$.  Correspondingly, the rotational transitions for linear molecules are spaced at even frequency intervals.  It is clear that molecules with a larger $B$ (i.e.,~lower moment of inertia; lighter or more compact molecules) have wider spacing between energy levels, and transitions at higher frequencies.  Similarly, molecules with a smaller B (i.e.,~higher moment of inertia; heavier or larger molecules) have more closely spaced transitions that fall at lower frequencies.

Spherical tops (e.g.~CH$_4$) have three identical moments of inertia, and exhibit similar energy spacing as linear molecules.  Symmetric tops have two identical moments of inertia and one axis with a different moment of inertia, and can be either prolate (cigar-shaped, e.g.~CH$_3$CN) or oblate (pancake-shaped, NH$_3$).  In this case, two rotational quantum numbers are needed to describe the rotational states: $J$ and $K$. $J$ is the rotational quantum number (as before), and $K$ is the quantum number associated with the projection of the rotational angular momentum onto the symmetry axis, with $K \leq J$. The selection rule for transitions is $\Delta J = \pm1$ and $\Delta K = 0$.  Because radiative transitions between $K$-levels are not allowed, the relative population of different $K$-levels is set purely by collisions, making symmetric tops excellent temperature probes.  Transitions with different $K$ quantum numbers but the same upper and lower $J$-levels occur close together in frequency, and are often referred to as `rotational ladders'. Finally, asymmetric tops are molecules for which all three orthogonal rotational axes have different moments of inertia.  For such molecules (e.g.~H$_2$O, CH$_3$OH) there is no simple equation representing the rotational energy levels. These molecules have chaotic and often very complex rotational spectra. Most molecules fall into this category.

Even small linear molecules may have a more complicated energy structures and spectra than described above due to nuclear spin effects. For nuclei with non-zero spin, interactions between the nuclei and the electrons result in so-called hyperfine splitting of the rotational energy levels. CN, HCN and C$^{17}$O are examples of molecules routinely observed in disks that display hyperfine structure in their spectra. For molecules that contain two or more hydrogen atoms, such as H$_2$O and H$_2$CO, the rotational transitions are split into two ladders for molecules with either the hydrogen nuclear spins aligned parallel (ortho) or anti-parallel (para). Radiative transitions between these ortho and para states are forbidden to high order, and only chemical reactions can transform one state into the other. The ortho-to-para ratio can take values between 0 and 3. Furthermore, transitions from molecules with unpaired electrons, such as CN, can be split into multiple components in the presence of a magnetic field (`Zeeman splitting').

\subsection{Infrared observations} \label{sec:Infrared}

\begin{figure}[h]
\centering
\includegraphics[width=0.8\textwidth]{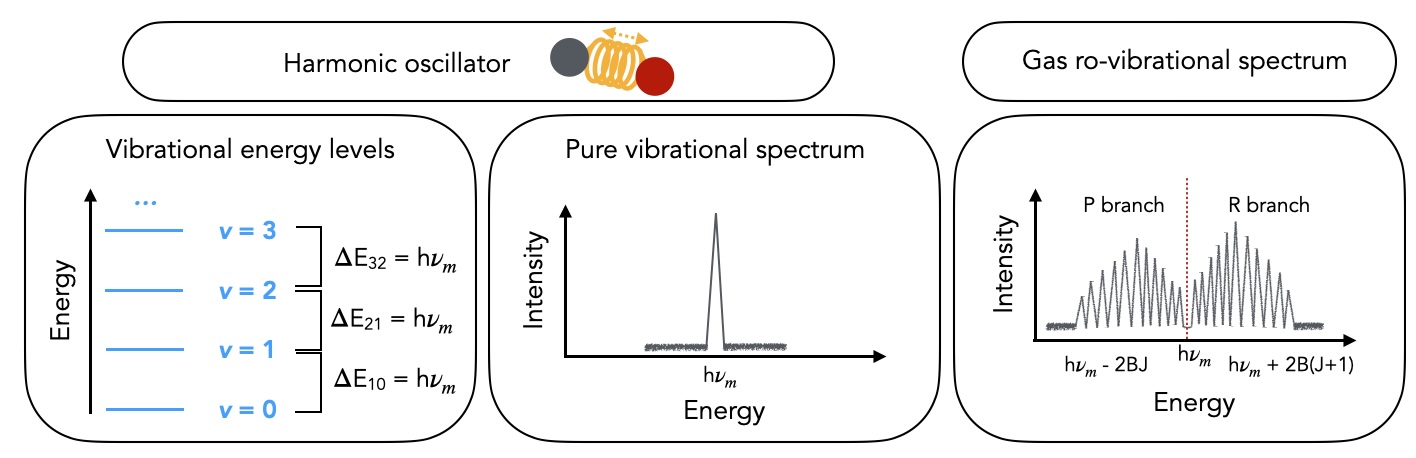}
\caption{Illustration of vibrational energy levels for a harmonic oscillator (left, middle panels).  Energy levels are evenly spaced, leading to a single vibrational transition energy with $\Delta E = h\nu_m$.  In reality, anharmonicity introduces weaker `overtone' features into a vibrational spectrum.  Also, vibrational transitions are often accompanied by rotational transitions, leading to a series of ro-vibrational transitions (right panel). The vertical red line marks the location where the Q branch would be if it were allowed. }
\label{fig:spectroscopy_vib}
\end{figure}

The advantage of infrared observations is that vibrational transitions can be observed for molecules both in the gas and in the ice, as well as for some molecules without a permanent dipole moment. Vibrational levels are denoted by the vibrational quantum number $\nu$.  The vibrational energy levels can be approximated by a harmonic oscillator (Figure \ref{fig:spectroscopy_vib}): $E_\nu = (\nu + \frac{1}{2})h\nu_m$, where $\nu_m$ is the fundamental frequency, proportional to the bond stiffness and inversely proportional to the reduced mass.  For a perfect harmonic oscillator, only transitions with $\Delta \nu = \pm 1$ are allowed.  This results in a transition energy spacing of $\Delta E = h \nu_m$, which is the same for all allowed transitions: an ideal harmonic oscillator would have a vibrational spectrum consisting of just one line at this frequency, called the `fundamental' transition.  Real molecules are not perfect harmonic oscillators, and weak `overtone' transitions are possible with $\Delta \nu > |1|$.  Anharmonicity also causes different fundamental transitions ($\Delta \nu = \pm 1$) to occur at different frequencies in real molecules.

In order for a molecular vibration to be IR-active, there must be a change in the dipole moment as a result of the vibration.  This means that vibrational transitions are allowed for symmetric polyatomic molecules (e.g.~CO$_2$, CH$_4$) which have no permanent dipole moment and therefore no rotational spectrum.  Vibrational transitions remain forbidden for homonuclear diatomics (e.g.~N$_2$, O$_2$) which have no bond dipole.  For a non-linear molecule with $N$ atoms, there are $3N-6$ different vibrational modes (e.g., different ways in which the molecule can bend and stretch).  Linear molecules have $3N-5$ different vibrations, because they have one less degree of freedom.  Not all modes lead to a distinct transition energy due to degeneracies and to IR-inactive modes.  

In the gas phase, vibrational transitions are often accompanied by rotational transitions with $\Delta J = 0, \pm 1$ (ro-vibrational transitions).  This leads to 3 distinct `branches' of a vibrational transition.   
The R-branch refers to transitions with $\Delta\nu = \pm 1$ and $\Delta J = +1$, and the P-branch to transitions with $\Delta\nu = \pm 1$ and $\Delta J = -1$.  For an ideal diatomic, the transition energies would be described $\Delta E_R (\nu,J) = h\nu_m + 2B(J+1)$ and $\Delta E_P (\nu,J) = h\nu_m -2BJ$.  Each line is labeled R($J$) or P($J$), where $J$ represents the value of the lower rotational state. The Q branch refers to transitions with $\Delta\nu = \pm 1$ and $\Delta J = 0$, and is forbidden in many cases. A molecule relevant to disks that does have an allowed Q-branch is CO$_2$. Because the rotational constant, $B$, is different in each vibrational state, the R-branch lines move closer together in the spectrum for increasing $J$, and eventually start moving closer to the Q-branch again. The line at the highest frequency is called the bandhead. The opposite happens for the P-branch, where the lines move further apart for increasing $J$. 

Vibrational transitions from solid-state molecules (including ices) do not have a rotational substructure and are very broad. The line profiles depend on the morphology (i.e., crystalline or amorphous ices), the environment of the observed molecule (i.e., what other molecules are present in the ice), and the temperature of the ice.  The band positions are shifted relative to the gas phase due to interactions with neighboring molecules.  Ices are best observed in absorption in the near- to mid-infrared, and can therefore only be seen toward an infrared-bright background source. For disks, this means that ices have to be observed in sources with an edge-on geometry.  A challenge with ice spectroscopy is the lack of specificity when identifying molecular carriers: it can be difficult to distinguish between different molecules that have the same functional group (e.g., -OH), because such a group has the same bending and/or stretching modes in different molecules. Identification is further complicated by the broad bands and environment-specific band position and profile.


\section{Analysis of molecular line emission from disks} \label{sec:Analysis}

\begin{figure}[!t]
\centering
\includegraphics[width=\textwidth]{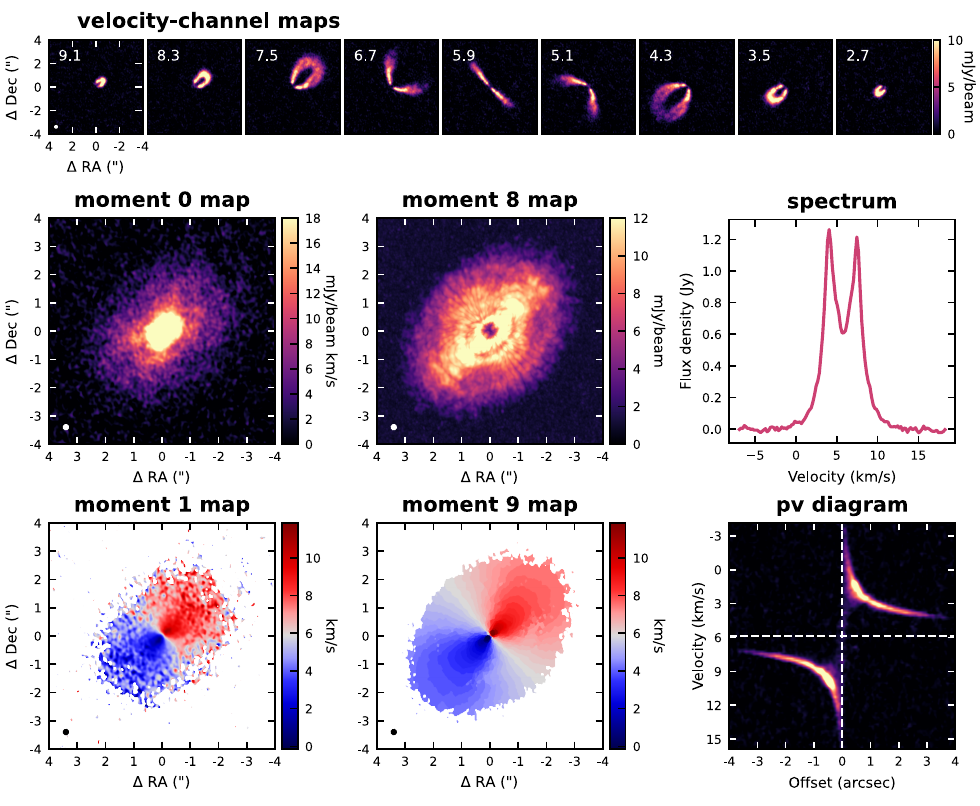}
\caption{Examples of different ways to display molecular line observations using C$^{18}$O observations of the HD 163296 disk from the ALMA Large Program MAPS \citep{Oberg2021}. Top row: Selected velocity-channel maps with the velocity (km s$^{-1}$) indicated in the top left corner. Middle row: Moment 0 map (velocity integrated intensity), moment 8 map (peak intensity), and spectrum or line profile (disk-integrated intensity). Bottom row: Moment 1 map (intensity-weighted velocity), moment 9 map (velocity at peak intensity) and position-velocity (pv) diagram along the disk major axis (i.e., from northwest to southeast). The colorscale of the pv diagram is the same as for the moment 8 map. The small ellipse in the bottom left corner of the velocity-channel maps and moment maps represents the synthesized beam size of the interferometric observations. }
\label{fig:momentmaps}
\end{figure}

There exist many ways to extract physical parameters (e.g.~column density, rotational temperature) from molecular line observations, and the appropriate analysis route depends on the observational characteristics (e.g., spatial resolution, or whether the lines are weak or strong) and the science goals.  Spatially resolved observations consist of three-dimensional data `cubes' with two position axes and one frequency axis.  These are commonly visualized in several ways (Fig.~\ref{fig:momentmaps}).  Spectra show the flux density, either in a single pixel or spatially integrated, as function of frequency or doppler-shifted velocity.  Channel maps show a series of two-dimensional spatial images, each imaged at different frequencies.  A `moment map' is a two-dimensional image representing some statistic of the 3D cube along the spectral axis. The most common moment maps are zeroth moment (velocity-integrated intensity per pixel), first moment (intensity-weighted velocity per pixel), eighth moment (peak intensity per pixel), and ninth moment (velocity at peak intensity per pixel). For young disks, position-velocity diagrams are often used to separate emission from the disk and envelope, by presenting the emission along the major axis of the disk as function of velocity. 

The signal-to-noise ratio of weak lines can be enhanced by utilizing the Keplerian velocity pattern of the disk. For example, a `Keplerian mask' including only pixels expected to contain emission from the disk can be used when creating moment maps \citep{Salinas2017}, and a `shift-and-stack` technique whereby the velocity of each pixel is shifted to the rest frequency can enhance signal-to-noise in line spectra \citep{Yen2016}. Finally, matched-filtering can be used to search for very weak emission lines using a strong line or model line profile as a template \citep{Loomis2018}.

Often, line emission is converted into a molecular column density ($N$) using the relation 
\begin{equation}\label{eq:coldens}
    \frac{4\pi F \Delta v}{A_{\rm{ul}}\Omega h c g_{\rm{u}}} = \frac{N}{Q(T)} e^{-E_{\rm{u}}/k_B T},
\end{equation}
where $F \Delta v$ is the velocity-integrated flux, $\Omega$ is the solid angle subtended by the source, $A_{\rm{ul}}$ is the Einstein A coefficient of the transition, $E_{\rm{u}}$ and $g_{\rm{u}}$ are the energy and degeneracy of the upper level, respectively, $T$ is the temperature of the emitting material, $Q(T)$ is the molecular partition function at temperature $T$, $h$ is the Planck constant and $c$ is the speed of light. For weak lines or spatially unresolved observations, the emission is typically integrated over the entire disk area resulting in 'disk-integrated column densities'. When the signal-to-noise is high enough, radial column density profiles may be extracted. 

The above calculation assumes that the material is in local thermodynamic equilibrium (LTE), that is, that the density is high enough for the population of the different energy levels to be determined by collisions. This is determined by a molecular transition's `critical density': the density, at a given temperature, at which the rate of collisional excitation to the upper level of the transition equals the rate of radiative de-excitation back into the lower level. If the density is below the critical density (non-LTE conditions), the balance between collisional excitation, collisional de-exciation and radiative de-excitation has to be calculated explicitly. Typically, LTE is a valid assumption when analyzing disk emission. The above relation (Eq.~\ref{eq:coldens}) also requires the emission to be optically thin. Observations of multiple isotopologues are often used to assess the optical depth of the emission as the flux ratio of the same transition for different isotopologues should equal the isotope ratio in case of optically thin emission. Finally, the temperature of the emitting material can be determined if multiple transitions of a given molecule are observed using the ratio of the different line intensities. In this case, temperature and column density are typically determined simultaneously. This can either be done in a `rotation diagram' analysis from a linear fit to $\log{(N_{\rm{u}}/g_{\rm{u}})}$ (which is equal to the left side of Eq.~\ref{eq:coldens}) versus $E_u$, or by fitting synthetic spectra for varying column density and temperature. If the temperature cannot be derived from the observations directly, an assumption has to be made based on other knowledge of the target. 

When LTE is not valid or when one wants to derive more detailed information about the spatial distribution of a molecule, radiative transfer models can be employed to produce synthetic images for comparison with the observations. Publicly available codes exist with different degrees of complexity. For example, \texttt{RADEX} deals with a one-dimensional or `slab' models \citep{vanderTak2007}, \texttt{Ratran} can be used for two-dimensional models (e.g., radial column density profiles; \citealt{Hogerheijde2000}), and \texttt{LIME} and \texttt{RADMC3D} can handle full three-dimensional geometries \citep[][respectively]{Brinch2010,Dullemond2012}. The input molecular distribution can also be created at different levels of complexity, ranging from simple constant abundances within varying physical regions (`parametrized' distributions) to outputs from full chemical networks. It may not come as a surprise that chemical models also vary in degrees of complexity, for example in terms of the size of the chemical network that is used, that is, the number of molecular species and chemical reactions that are included. In addition, chemical models can be separated into pure chemical models, where the chemical structure is calculated for a given input physical structure, and thermo-chemical models where the gas temperature and chemical structure are calculated self-consistently; because the chemical structure depends on the temperature and the temperature is influenced by atomic and molecular line cooling, these models incorporate an iterative process of calculating the temperature and chemical structure.


\section{Future outlook} \label{sec:Outlook}

While ALMA has been operational for a little over a decade, it has not yet reached its full potential regarding studies of disk chemistry. Deep high-resolution observations of individual or small groups of disks in combination with newly developed analysis techniques are likely to reveal the chemical structure, potentially at the site where planets are forming, in unprecedented detail. Another promising avenue are `population' studies of large numbers of disks that are now possible through Large Programs. While not (yet) as detailed as studies of individual disks, such programs will reveal global chemical properties of disks and how they change with disk evolution, as well as how chemically diverse disks really are. JWST has only just started operations and the initial results for inner disk chemistry and ice composition, as well as exoplanet atmospheres, are only just scratching the surface. At the same time, the development of more advanced chemical models incorporating more and more details of disk physics, grain growth and planet formation will provide new predictions to be tested observationally. This combination of state-of-the art facilities (ALMA, JWST, and future ground-based Extremely Large Telescopes) and models is bound to significantly increase our understanding of planet composition in the next decade.

\begin{ack}[Acknowledgments]
WWe would like to thank Lenore Anderson, Ted Bergin, Deniz Ka\c{c}an, Morgan Kennebeck, Vincent Kreft, Patricia Moon, William Thompson, Anna Wannenmacher, and Lauren Warshaw for valuable feedback on the manuscript. 
\end{ack}

\seealso{Recent in depth reviews on disk chemistry are provided by \citet{Aikawa2022} and \citet{Oberg2023}. }

\bibliographystyle{Harvard}
\bibliography{reference}

\end{document}